\documentclass[journal]{IEEEtran}
\usepackage{cite}
\ifCLASSINFOpdf
  \usepackage[pdftex]{graphicx}
\else
  \usepackage[dvips]{graphicx}
\fi
\usepackage{amsmath}
\usepackage{amssymb}

\usepackage{array}
\ifCLASSOPTIONcompsoc
  \usepackage[caption=false,font=normalsize,labelfont=sf,textfont=sf]{subfig}
\else
  \usepackage[caption=false,font=footnotesize]{subfig}
\fi
\usepackage{url}
\usepackage{enumerate}
\usepackage{color}

\begin{document}
\title{A Reciprocal Heuristic Model for Diffuse Scattering from Walls and Surfaces}

\author{Enrico~M.Vitucci,
        Nicolò~Cenni,
        Franco~Fuschini,
        and~Vittorio~Degli-Esposti% <-this % stops a space
\thanks{All the Authors are with the Department
of Electrical, Electronic, and Information Engineering "Guglielmo Marconi" (DEI), CNIT University of Bologna, 40126 Bologna, Italy.}% <-this % stops a space
\thanks{Corresponding author: Enrico M. Vitucci (e-mail: enricomaria.vitucci@unibo.it)} }

\maketitle

\begin{abstract}
Diffuse scattering of electromagnetic waves from natural and artificial surfaces has been extensively studied in various disciplines, including radio wave propagation, and several diffuse scattering models based on different approaches have been proposed over the years, two of the most popular ones being Kirchhoff Theory and the so-called Effective Roughness heuristic model. The latter, although less rigorous than the former, is more flexible and applicable to a wider range of real-world cases, including non-Gaussian surfaces, surfaces with electrically small correlation lengths and scattering from material inhomogeneities that are often present under the surface. Unfortunately, the Effective Roughness model, with the exception of its Lambertian version, does not satisfy reciprocity, which is an important physical-soundness requirement for any propagation model.
In the present work, without compromising its effectiveness and its simple and yet sound power-balance approach, we propose a reciprocal version of the Effective Roughness model, which can be easily implemented and replaced to the old version in ray-based propagation models. The new model is analyzed and compared to the old one and to other popular models. Once properly calibrated, it is shown to yield similar - if not better - performance with respect to the old one when checked vs. measurements.
\end{abstract}

% Note that keywords are not normally used for peer review papers.

% \fi
\IEEEpeerreviewmaketitle

\section{Introduction}

Diffuse scattering (DS) of radio waves, intended here as non-specular reflection from terrain, objects and building walls surfaces due to surface roughness or material irregularities, has been studied for years in many application fields such as remote sensing and optics.

With reference to radio propagation in urban environment, assuming flat, smooth and homogeneous building walls or indoor furniture panels, propagation can be conveniently analyzed using the Geometrical Optics (GO) approximation \cite{Felsen1973}, where radio wave interactions can be modeled as specular reflections, transmissions and edge diffractions.

However, perfectly smooth slabs are rarely present in real-life, especially in dense urban areas where building walls can show relevant deviations from smooth homogeneous layers, such as compound materials, windows frames, metal reinforcements, pillars, rough plaster and brick surfaces, cables, advertising boards, etc. Similar considerations hold true for indoor walls and furniture. In fact, some investigations showed that DS due to such details - often disregarded in building maps and databases - can be an important propagation mechanism in urban environment \cite{DEB1999,VDEAP2001,VDEAP2007,vitucci2018scalemodel}. In particular, DS has been shown to generate a large part of the time-domain, angle-domain and polarization dispersion of the multipath radio channel in most environments \cite{VDEAP2004,Mani2012,vitucci2012polarimetric,mani2014parameterization}, and the knowledge of this phenomena can be exploited in the design of MIMO wireless links and to implement advanced beamforming strategies\cite{vitucci2014MIMO,VDE2014RTbeamforming,fuschini2019studybeamforming}. Moreover, DS has been shown to play a prominent role even in the determination of the actual RF coverage level, especially in Non Line of Sight (NLoS), millimeter-wave frequency applications \cite{Tengfei2019}. Recent studies have also highlighted the importance of DS from rough surfaces in Terahertz wireless communications links\cite{SheikhMIMOTHz,XieDiffuseTHz2022,JansenTHz2011}. Therefore, accounting for specular reflection, transmission and diffraction is not sufficient: analysis and modeling of diffuse scattering is mandatory to achieve a complete understanding of urban radio propagation.

The most widely known diffuse scattering models available in the literature only deal with surface roughness, and include Kirchhoff Theory, the Small Perturbation Method and the Integral Equation Method \cite{BeckmannSpizzichino63, Tsang2000}. The most popular approach to DS is Kirchhoff Theory, based on Beckmann-Kirchhoff theory for scattering of incident plane waves from Gaussian rough surfaces described in terms of roughness’ standard deviation and correlation distance \cite{BeckmannSpizzichino63}.
Another diffuse scattering approach developed specifically for building walls and derived from physical optics is proposed in \cite{Bertoni2004}: here the assumption is that non-specular scattering from the façades of large buildings is dominated by windows and decorative masonry, whose placement tends to be nearly periodic.

However, Kirchhoff Theory is not applicable to non-Gaussian surface roughness, to strong surface irregularities where the roughess’ correlation length is comparable to, or smaller than, the wavelength (e.g. indentations), or when the surface size is comparable to, or smaller than, correlation length. Moreover all the cited models are not suitable to cases where the presence of internal, material irregularities have a significant impact. The possibility for radiowaves to penetrate inside the wall, undergo scattering interactions due to the internal inhomogeneities and re-emerge with nearly random propagation direction and characteristics must also be accounted for.

Therefore, in more recent years heuristic models like the Effective Roughness (ER) model \cite{VDEAP2007} have been proposed to overcome the foregoing limitations. The ER model is aimed at modeling non-specular scattering from surfaces, but its parameters are not actual surface roughness parameters as for the Kirchhoff model, but "effective" parameters that must take into account also the effect of the more general irregularities and details described above, hence the name "Effective Roughness" model. Differently from the Kirchhoff model, the specular reflected wave and the scattered wave are treated from the beginning as distinct waves where the attenuation of the former is due to part of its power being diverted into the latter by irregularities. This fact allows its straightforward, “plug-and-play” integration into ray-based models where specular reflection and transmission are implemented as phase-coherent waves that follow GO theory, albeit with a proper attenuation, while diffuse scattering can have different spatial and polarization characteristics.

The ER model is physically consistent as it is based on a power balance between specular reflection, transmission and scattering. It is flexible because the scattering pattern can be chosen among several different options, and due to its simplicity and low number of parameters it can be easily tuned vs. measurement data.

After its introduction in 2007, analytical formulations of the ER model have been developed to describe the angle-spread produced by DS from a single wall \cite{DeFuVi2009}, it has been extended to transmitted scattering in the forward half-space (e.g. beyond a wall) \cite{FuDeVi2010} and has been further validated vs. full-wave electromagnetic simulations and measurements in reference cases \cite{MiDeDe2014}. The parameterization of the ER model in the mm-wave bands for different construction materials has also been discussed in \cite{Pascual2016,ItemLevel2016}.
Furthermore, the ER model has been finally embedded into some commercial ray based field prediction software tools \cite{Remcom}.

Despite its strengths, the original - or legacy - ER model also has an important shortcoming: with the exception of its Lambertian scattering pattern version, it doesn't fully satisfy reciprocity, which means that the predicted scattered field intensity is not invariant with respect to the exchange of transmitter and receiver, as it should be according to propagation theory \cite{VanBladel07}. Although, being a heuristic model, its fitting to the actual physical process can always be adjusted through parameter calibration, non-reciprocity represents an important theoretical flaw, especially considering that its non-reciprocal, directive scattering versions have been shown to be the most suitable to describe DS from real buildings \cite{VDEAP2007}.

Other models similar to the ER model that satisfy reciprocity have been developed for computer graphic applications \cite{Duvenhage2013,WaMaHo2007}, or have been derived from them \cite{Wagen2020}. However, such models do not distinguish specular from diffuse reflection and therefore cannot be easily implemented into existing ray-based propagation models. Moreover, although power constraints are present, such as that the back-scattered power cannot be greater than the incident power, they don’t comply with a clear power conservation balance at the surface in order to minimize parameters and to achieve maximum compatibility with traditional formulations based on Geometrical Optics for smooth surfaces and material slabs.

In the present work, starting from the approach of the original ER model, we first develop a better and more complete mathematical derivation of its normalization factors with respect to the rather incomplete demonstration provided in \cite{VDEAP2007}, using Euler's Gamma and Beta functions. Then we propose a new version of the ER model that satisfies reciprocity without sacrificing the original power-balance assumptions, if not to a negligible extent for grazing incidence. We also provide a discussion on reciprocity and power balance of the new ER model with respect to the original formulation, and a comparison with respect to other reference models (e.g. Kirchhoff). Finally, the model is validated through comparison with measurements in a reference case.

The paper is organized as follows. In Section II, some background on the original ER model and its formulation are provided, then the new reciprocal formulation is presented (the mathematical details are provided in the appendices). In Section III, comparisons to the legacy ER model, to other reference models and to measurements are shown and discussed. Finally, conclusions are drawn in Section IV.

\section{The New ER Model's Formulation}
\subsection{Background on the ER Model}
\begin{figure}[!ht]
  \centering
  \includegraphics[width=0.45\textwidth]{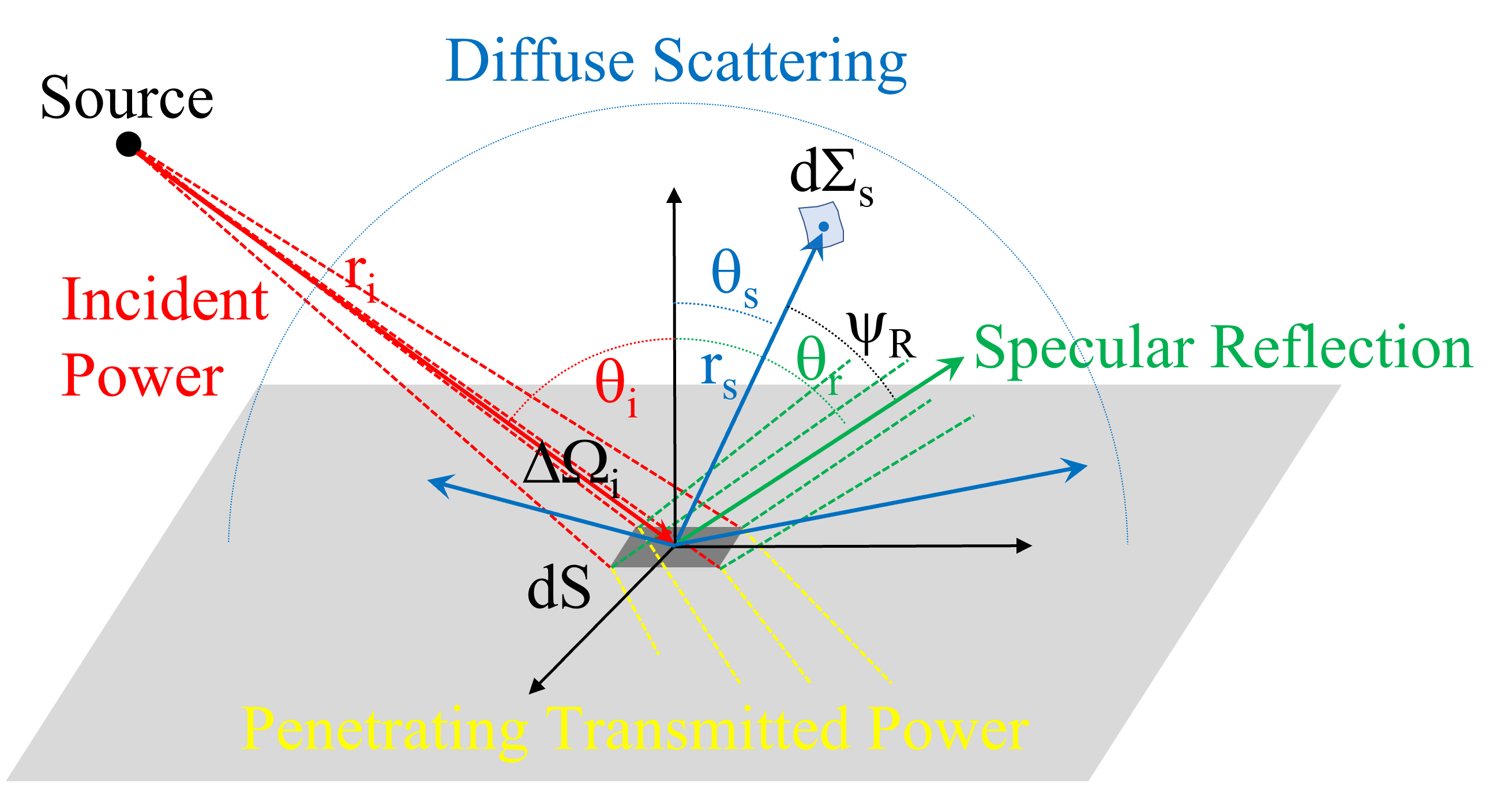}
  \caption{Power balance on a surface element: incident wave, specular reflection, diffuse scattering, and penetrated power.}
  \label{Fig_ER_model}
\end{figure}

When a surface element $dS$ is illuminated by an impinging electromagnetic wave, the following power balance must hold:
\begin{equation}
P_i=P_r+P_s+P_p
\label{eq:1}
\end{equation}
being $P_i$, $P_r$, $P_s$ and $P_p$ the
incident, the reflected, the scattered and transmitted powers, respectively (Fig. \ref{Fig_ER_model}).

The basic assumption of the ER approach is that the scattered power can be simply related to a scattering coefficient $S \in [0,1]$ as:
\begin{equation}
P_s=S^2 \cdot \left( U^2 P_i \right)
\label{eq:2}
\end{equation}
Depending on the value of $U$, $S^2$ represents the percentage of either the incident ($U=1$) or the reflected power ($U=\Gamma$, being $\Gamma=\left| \bar{E}_r \right| / \left| \bar{E}_i \right| $ the modulus of the reflection coefficient) that is spread in non specular directions \cite{VDEAP2007}. In the following, DS is supposed to occur at the expense of specular reflection, i.e. $U=\Gamma$ is considered in eq. (2).

Therefore, the power balance (\ref{eq:1}) can be written as:
\begin{equation}
1={\Gamma}^{2}R^2+{\Gamma}^{2}S^2+P_p/P_i
\label{eq:3}
\end{equation}
where $R$ is the reflection reduction factor, which is related to the so called "Rayleigh's factor" of Kirchhoff theory \cite{BeckmannSpizzichino63}.
By assuming that the ratio $P_p/P_i$ does not depend on the degree of roughness, i.e. on the parameter $S$, from (\ref{eq:3}) we easily get that the reflection reduction factor is \cite{VDEAP2007}: $R=\sqrt{1-S^2}$.

Power balance assumptions of the legacy ER model (referred to as \textit{ER power-balance} in the following), represented by eq. (\ref{eq:2}) and (\ref{eq:3}), imply the following equation where power diverted from specular reflection equals the integral of the scattered field power density over the backscattering half space, i.e. (Fig. \ref{Fig_ER_model})
\begin{equation}
\begin{split}
P_s&=S^2\Gamma^2 P_i=S^2\Gamma^2\frac{\left |\bar{E}_i  \right |^2}{2\eta}\Delta\Omega_i\:r_i^2=\\
&=\int_{0}^{2\pi}\int_{0}^{\frac{\pi}{2}}\frac{\left |\bar{E}_S \right |^2}{2\eta}r_S^2\: \sin\theta_S\:d\theta_S\,d\phi_S
\end{split}
\label{eq:4}
\end{equation}
where $r_i$, $r_S$ are the distances between the surface element $dS$ and the source, and between $dS$ and the observation point, respectively, $\Delta\Omega_i$ is the solid angle subtended from $dS$ at the transmitter side (see Fig.\ref{Fig_ER_model}), and $\eta=\sqrt{\mu_0/\epsilon_0}$ is the free-space impedance.

Moreover, the squared amplitude of the scattered field is assumed to be expressed by the following formula:
\begin{equation}
\left | \bar{E}_S \left(\theta_i,\phi_i,\theta_S,\phi_S\right)\right |^2=E_{S0}^2 \cdot \,f(\theta_i,\phi_i,\theta_S,\phi_S)
\label{eq:5}
\end{equation}
where $f\left(\theta_i,\phi_i,\theta_S,\phi_S\right) \in \left[0,1\right]$ represents the diffuse scattering spatial pattern. The overall scattered field is then modelled as a non-uniform spherical wave.

Assuming the surface element $dS$ in the far field region of the transmitting source, the incident field is a spherical wave, and therefore:
\begin{equation}
\left |\bar{E}_i\left(r_i,\theta_i,\phi_i \right)  \right | = \frac{\sqrt{60\;P_t\;g_t\left(\theta_i,\phi_i\right)}}{r_i} = \frac{K_i\left(\theta_i,\phi_i\right)}{r_i}
\label{eq:6}
\end{equation}
where $K_i\left(\theta_i,\phi_i\right)$ is a parameter depending on the source properties (transmit power, antenna gain). By substituting (\ref{eq:5}) and (\ref{eq:6}) into (\ref{eq:4}) and exploiting the expression for the solid angle $\Delta\Omega_i=\frac{dS\,\cos\theta_i}{r_i^2}$, the following formula can be achieved:
\begin{equation}
\left |\bar{E}_S \right |^2=\underbrace{\frac{\left ( \frac{K_i S}{r_i\,r_S} \right )^2 \Gamma^2\,dS\,\cos\theta_i}{F\left(\theta_i,\phi_i\right)}}_{E_{S0}^2} \cdot f\left(\theta_i,\phi_i,\theta_S,\phi_S\right)
\label{eq:7}
\end{equation}
where $F\left(\theta_i,\phi_i\right)$ represents the following integral expression:
\begin{equation}
F\left(\theta_i,\phi_i\right)=\int_{0}^{2\pi}\int_{0}^{\frac{\pi}{2}} f(\theta_i,\phi_i,\theta_S,\phi_S)sin\,\theta_S\:d\theta_S\,d \phi_S
\label{eq:8}
\end{equation}
It can be observed that, according to (\ref{eq:7}), $\left|\bar{E}_S\right|=0$ for any observation angle, when the incident wave is parallel to the surface element, i.e. $\theta_i=\pi/2$: in fact, for grazing incidence, no power is captured and then scattered by the surface.

It is worth noting that, in order to have a reciprocal expression for the intensity of the scattered field $\left | \bar{E}_S \right |$, the product of the three functions in (\ref{eq:7}) needs to be reciprocal, i.e.:
\begin{equation}
\frac{f\left(\theta_i,\phi_i,\theta_S,\phi_S\right)}{F(\theta_i,\phi_i)} \cdot cos\,\theta_i = g_{rec}\left(\theta_i,\phi_i,\theta_S,\phi_S\right)
\label{eq:9}
\end{equation}
where $g_{rec}$ is a reciprocal function, i.e. a function invariant to the exchange of $\left(\theta_i,\phi_i\right)$ with $\left(\theta_S,\phi_S\right)$.

The former version of the scattering model in \cite{VDEAP2007} was aimed at a single-lobe, directive scattering pattern by means of the following choice:
\begin{equation}
f\left(\theta_i,\phi_i,\theta_S,\phi_S\right)=\left ( \frac{1+\cos\psi_R}{2} \right )^{\alpha_R}\;\;\;\alpha_R \in \mathbb{N}
\label{eq:10}
\end{equation}
where the exponent $\alpha_R$ is a tuning parameter for the directivity of the scattering pattern (the greater $\alpha_R$, the narrower the lobe), and $\psi_R$ is the angle between the scattering direction $\left(\theta_S,\phi_S\right)$ and the specular direction (Fig. \ref{Fig_ER_model}). The following relation is also provided in \cite{VDEAP2007}:
\begin{equation}
\cos\psi_R=\cos\theta_i\cos\theta_S-\sin\theta_i\sin\theta_S\cos\left ( \phi_S-\phi_i \right )
\label{eq:11}
\end{equation}
By applying the power balance (\ref{eq:4}), equation (\ref{eq:7}) becomes:
\begin{equation}
\left | \bar{E}_S \right |^2=\left (\frac{K_i S}{r_i \, r_S}  \right )^2 \Gamma^2\,\frac{dS\,\cos\theta_i}{F_{\alpha_R}\left(\theta_i\right)}\left ( \frac{1+\cos\,\psi_R}{2} \right )^{\alpha_R}
\label{eq:12}
\end{equation}
where $F_{\alpha_R}$ is the solution of the integral in (\ref{eq:8}), when (\ref{eq:10}) is enforced \cite{VDEAP2007}. Note that with the chosen shape for the scattering pattern in eq. (\ref{eq:10}), $F_{\alpha_R}$ does not depend on the azimuth angle $\phi_i$, for symmetry reasons.

A complete solution for $F_{\alpha_R}$ was not derived in \cite{VDEAP2007}: however, two different, closed-form expressions were proposed, depending on whether $\alpha_R$ is even or odd.

Instead, a more compact and general expression for $F_{\alpha_R}$ is fully derived in this work, by exploiting the properties of the Euler's Beta function (Appendix A). The new closed-form solution valid for any value of $\alpha_R$ is (see Appendix B):
\begin{equation}
\begin{split}
F_{\alpha_R}\left(\theta_i\right)&=\frac{2\pi\,\alpha_R !}{2^{\alpha_R}}\sum_{j=0}^{\alpha_R}\frac{1}{\left ( \alpha_R-j \right )!\left ( j+1 \right )!!}\cdot\\
&\cdot\sum_{l=0}^{\left \lfloor j/2 \right \rfloor}\frac{\cos^{j-2l}\theta_i\sin^{2l}\theta_i}{2^{l}\,l!\left ( j-2l \right )!!} \:\:\:\:\:\:\:\:\:0\leq\theta_i<\frac{\pi}{2}
\end{split}
\label{eq:13}
\end{equation}
where $\lfloor x \rfloor$ stands for the greatest integer less than or equal to $x$, and the $!$ and $!!$ symbols stand for the factorial and double factorial functions, respectively (see Appendix A).
In the case of normal incidence ($\theta_i=0$), (\ref{eq:13}) reduces to:
\begin{equation*}
F_{\alpha_R}\left(0\right)=\frac{4\pi}{\alpha_R+1}\left(1-\frac{1}{2^{\alpha_R+1}}\right)
\end{equation*}

Looking at equations (\ref{eq:12}) and (\ref{eq:13}), it is evident that the amplitude of the scattered field $\left | \bar{E}_S \right |^2$ is non reciprocal, due to the presence in (\ref{eq:13}) of the functions $(\cos\theta_i)^m$ and $(\sin\theta_i)^m$, which are not counterbalanced by similar terms containing the scattering elevation angle $\theta_S$.

\subsection{The new reciprocal formulation}
The aim is to achieve a reciprocal expression for the scattered field. To this extent, we propose the following new expression for the diffuse scattering pattern:
\begin{equation}
f\left( \theta _i,\phi _i,\theta _S,\phi_S\right )=\sqrt{\cos\theta_S}\left (\frac{1+\cos\psi_R}{2}\right )^{\alpha_R}\:\:\:\alpha_R \ge 0
\label{eq:14}
\end{equation}

Such scattering function is obtained by multiplying the pattern of the legacy ER model, i.e. (\ref{eq:10}), by the factor $\sqrt{\cos\theta_S}$. With this assumption, the scattered power tends to zero for grazing observation angles, i.e. when $\theta_S$ approaches $\pi/2$. This is a necessary condition for reciprocity: in fact, according to (\ref{eq:7}), $\left|\bar{E}_S\right|=0$ for $\theta_i=\pi/2$, no matter what is the observation angle $\theta_S$ as the solid angle $\Delta\Omega_i$ goes to zero; similarly, it must be $\left|\bar{E}_S\right|=0$ for $\theta_S=\pi/2$, independently from the incidence angle $\theta_i$.

Besides, it can be observed that the multiplication of (\ref{eq:10}) by $\sqrt{\cos\theta_S}$ causes a skew of the maximum of the scattering pattern with respect to specular reflection. This disaligment is of the order of a few degrees, and is more evident for grazing incidence angles, and low values of the parameter $\alpha_R$: some examples will be shown and discussed in Section III.

Let's now discuss more in detail the reciprocity of the new model's formulation. In order for the model to be reciprocal, according to eq. (\ref{eq:9}) the following condition must be satisfied:
\begin{equation}
\frac{\cos\theta_i}{F_{\alpha_R}} \propto \sqrt{\cos\theta_i}
\label{eq:15}
\end{equation}
which implies: $F_{\alpha_R} \propto \sqrt{\cos\theta_i}$.

\begin{figure}[!ht]
  \centering
  \includegraphics[width=0.45\textwidth]{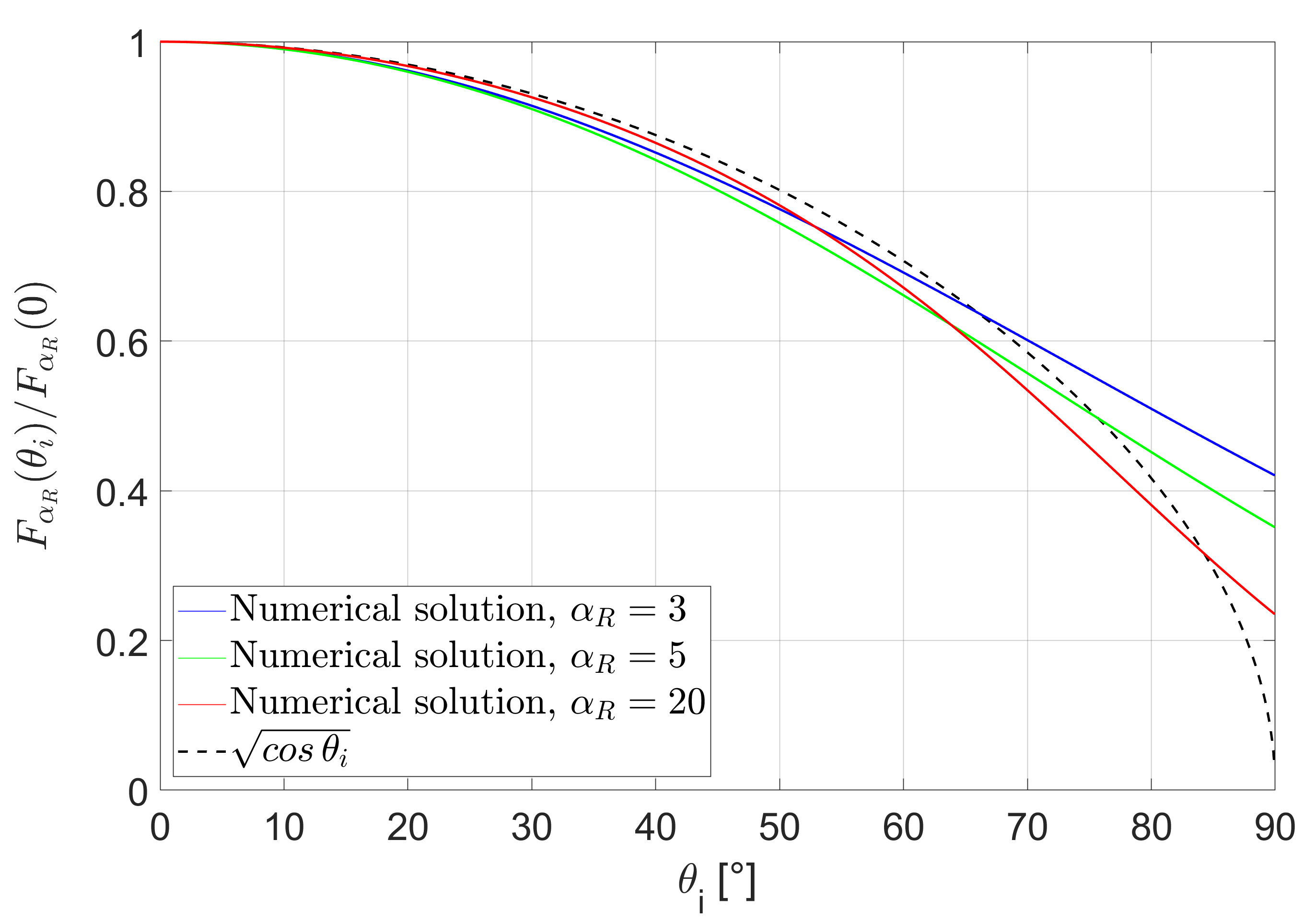}
  \caption{Comparison of the function $\sqrt{\cos\,\theta_i}$ with $F_{\alpha_R}\left(\theta_i\right)$, for different values of the exponent $\alpha_R$.}
  \label{fig:F_alpha_approx}
\end{figure}

Actually, it can be observed that $F_{\alpha_R}$ that would result from (\ref{eq:14}) being inserted into (\ref{eq:8}), i.e. which satisfies \textit{ER power-balance}, is a monotonic decreasing function having its maximum value for $\theta_i=0$, that can be well approximated by a function proportional to $\sqrt{\cos(\theta_i)}$, as shown in Fig. \ref{fig:F_alpha_approx}. This means that, if we assume $F_{\alpha_R} \propto \sqrt{\cos(\theta_i)}$, reciprocity is strictly satisfied, while also \textit{ER power-balance} is satisfied to a good extent.

In fact,  Fig. \ref{fig:F_alpha_approx} shows $F_{\alpha_R}\left(\theta_i\right)/F_{\alpha_R}\left(0\right)$ derived from (\ref{eq:8}) through numerical integration for 3 different values of the parameter $\alpha_R$, vs. the function $\sqrt{\cos\theta_i}$. It can be observed that the approximation is very good except for very grazing incidence angles (e.g. greater than $85^\circ$). This allows to write $F_{\alpha_R}$ in the form:
\begin{equation}
F_{\alpha_R}\left(\theta_i\right) \approx k\left ( \alpha_R \right )\sqrt{\cos\theta_i}
\label{eq:16}
\end{equation}
where $k\left ( \alpha_R \right )$ is an amplitude parameter depending only on the exponent $\alpha_R$. This approximation satisfies both eq. (\ref{eq:9}) (i.e. reciprocity) and, with good approximation, eq. (\ref{eq:8}) (i.e. \textit{ER power-balance}).

The value of $k\left ( \alpha_R \right )$ can be determined in a straightforward way by assuming the approximation (\ref{eq:16}) as valid and solving the integral (\ref{eq:8}) for $\theta_i=0$, as shown in Appendix C.

Then, the final reciprocal expression of the scattered field when (\ref{eq:14}) is enforced and under the approximation \eqref{eq:16} is (see Appendix C):
\begin{equation}
\setlength{\jot}{10pt}
\begin{split}
\left |\bar{E}_S \right |^2=&\left ( \frac{K_i S}{r_i\,r_S} \right )^2\Gamma^2 \, \frac{dS}{k\left(\alpha_R\right)} \, \cdot\\
&\cdot\sqrt{\cos\theta_i \cos\theta_S}\left ( \frac{1+\cos\psi_R}{2}\right)^{\alpha_R}
\end{split}
\label{eq:17}
\end{equation}
with
\begin{equation*}
k\left(\alpha_R\right)=\frac{4\pi}{2^{\alpha_R}} \sum\limits_{j=0}^{\alpha_R}\binom{\alpha_R}{j}\frac{1}{2j+3}
\end{equation*}
The expression used in \eqref{eq:17} for $k\left(\alpha_R\right)$ is valid only for integer positive values of the exponent $\alpha_R$, as it has been computed by using the binomial theorem (see Appendix C). However, in case real positive values of the exponent $\alpha_R$ are needed for a finer tuning of the model, $k\left(\alpha_R\right)$ can be calculated using the following interpolating function:
\begin{equation*}
k\left(\alpha_R\right) \approx
\left\{
\begin{aligned}
&(0.07937\alpha_R+0.1745)^{-1} && \text{if } \alpha_R>4\\
&(0.003128\alpha_R^2+0.05675\alpha_R+\\
&+0.2387)^{-1} && \text{if } 0 \le \alpha_R \le 4
\end{aligned}
\right.
\end{equation*}

\subsection{Double lobe model, reciprocal formulation}

\begin{figure}[!ht]
  \centering
  \includegraphics[width=0.25\textwidth]{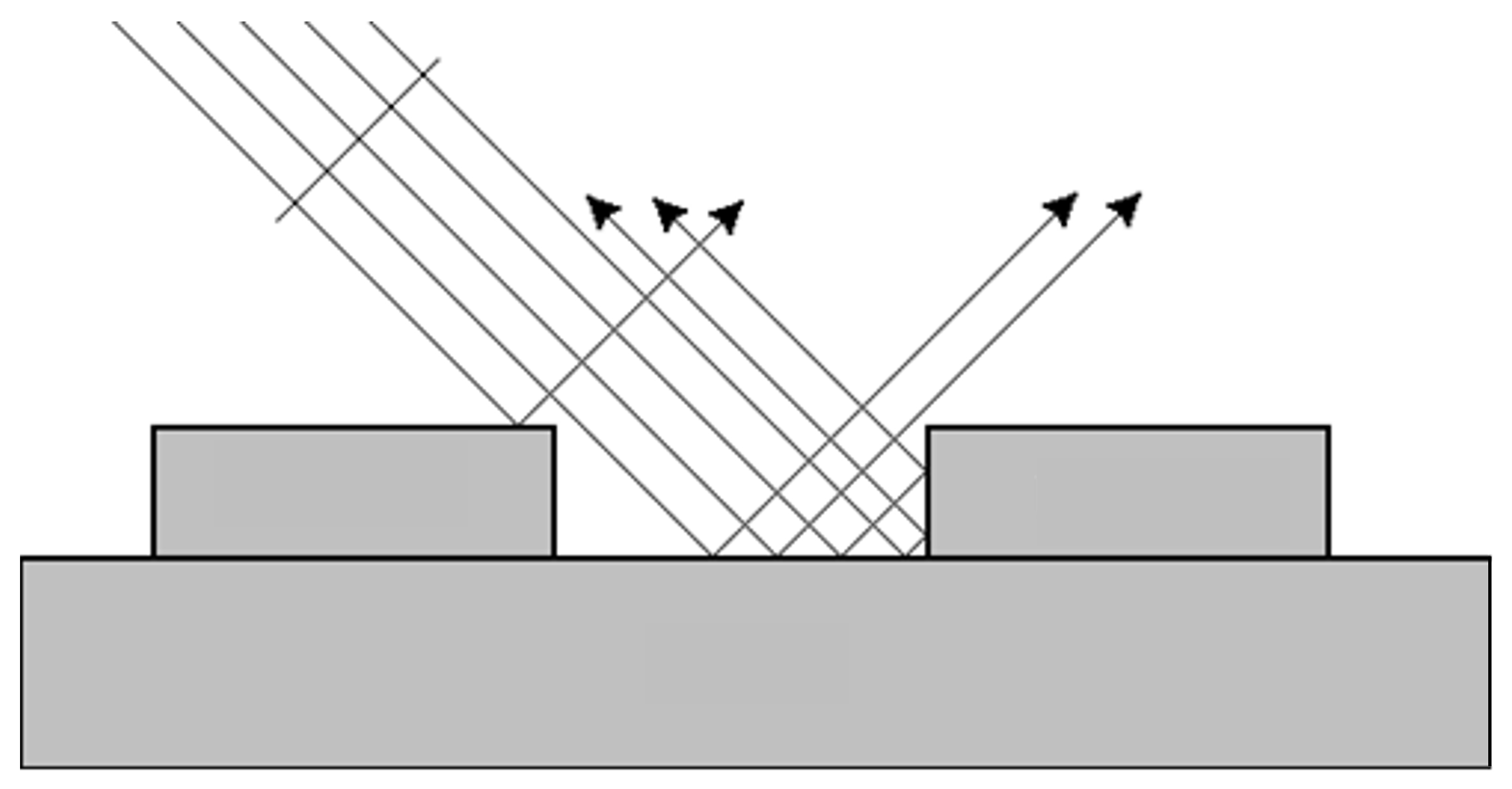}
  \caption{Wall with indentations that generate a backscattering component in the incidence direction}
  \label{fig:indentations}
\end{figure}

Similarly to what done in \cite{VDEAP2007} for the legacy ER model, it is possible to derive a double-lobe model, where an additional lobe steered in the incidence direction is added to the scattering pattern.
This is useful in many practical cases, where walls with big irregularities, e.g. indentations, generate a strong backscattering component in the incidence direction trough micro interactions consisting of multiple-bounce reflections (see Fig. \ref{fig:indentations}).

In order to obtain a reciprocal formulation for this double-lobe model, we propose the following expression for the scattered field

\begin{equation}
\setlength{\jot}{10pt}
\begin{split}
&\left |\bar{E}_S \right|^2=\left ( \frac{K_i S}{r_i\,r_S} \right )^2\,\Gamma^2\,\frac{dS\cos\theta_i}{F_{\alpha_i,\alpha_R}} \sqrt{\cos\theta_S} \:\cdot\\
&\:\:\:\:\:\cdot\left[\Lambda \left (\frac{1+\cos\psi_R}{2}\right )^{\alpha_R} + \left ( 1-\Lambda  \right )\left (\frac{1+\cos\psi_i}{2}\right )^{\alpha_i}\right]
\end{split}
\label{eq:18}
\end{equation}
with
\begin{equation}
\setlength{\jot}{10pt}
\begin{gathered}
F_{\alpha_i,\alpha_R}=\Lambda F_{\alpha_R} + \left( 1 - \Lambda \right) F_{\alpha_i}\\
F_{\alpha_R}=\int\limits_{0}^{2\pi}\int\limits_{0}^{\frac{\pi}{2}} \sqrt{\cos\theta_S} \, \left (\frac{1+\cos\psi_R}{2}\right )^{\alpha_R}     \sin\theta_S d \theta_S d \phi_S\\
F_{\alpha_i}=\int\limits_{0}^{2\pi}\int\limits_{0}^{\frac{\pi}{2}} \sqrt{\cos\theta_S} \, \left (\frac{1+\cos\psi_i}{2}\right )^{\alpha_i} \sin\theta_S d\theta_S d \phi_S
\end{gathered}
\label{eq:19}
\end{equation}

where $\psi_i$ is the angle formed by the observation and incidence directions, $\alpha_i$ is a parameter that accounts for the directivity of the backscattering lobe, $F_{\alpha_i,\alpha_R}$ is the solution of the power balance integral (\ref{eq:8}) for the double-lobe pattern, and $\Lambda \in \left[0,1\right]$ is a factor taking into account how the scattered power is subdivided between the two lobes.

It can be easily shown (see Appendix D) that both the integrals in (\ref{eq:19}) can be approximated by a function proportional to $\sqrt{\cos\,\theta_i}$, i.e. $F_{\alpha_R} \approx k\left ( \alpha_R \right)\sqrt{\cos\theta_i}$ and $F_{\alpha_i} \approx k\left ( \alpha_i \right)\sqrt{\cos\theta_i}$. If so, also $F_{\alpha_i,\alpha_R}$ is proportional to $\sqrt{\cos\,\theta_i}$, and this allows to get a reciprocal expression for the scattered field.
The final (reciprocal) expression of the scattered field for the double-lobe ER model is then:

\begin{equation}
\setlength{\jot}{10pt}
\begin{split}
\left |\bar{E}_S \right |^2=&\left ( \frac{K_i S}{r_i\,r_S} \right )^2\Gamma^2\,\frac{dS}{4\pi} \, \sqrt{\cos\theta_i\,\cos\theta_S}\: \cdot \\&\cdot \left[\Lambda\, \frac{2^{\alpha_R}}{\sum\limits_{j=0}^{\alpha_R}\binom{\alpha_R}{j}\frac{1}{2j+3}} \left (\frac{1+\cos\psi_R}{2}\right )^{\alpha_R} + \right. \\&
\left. + \left(1-\Lambda\right)\, \frac{2^{\alpha_i}}{\sum\limits_{j=0}^{\alpha_i}\binom{\alpha_i}{j}\frac{1}{2j+3}} \left (\frac{1+\cos\psi_i}{2}\right )^{\alpha_i} \right]
\end{split}
\label{eq:20}
\end{equation}

Note also that (\ref{eq:17}) is obtained as a particular case of (\ref{eq:20}), when $\Lambda$=1.

\begin{figure*}[!ht]
  \centering
  \includegraphics[width=0.8\textwidth]{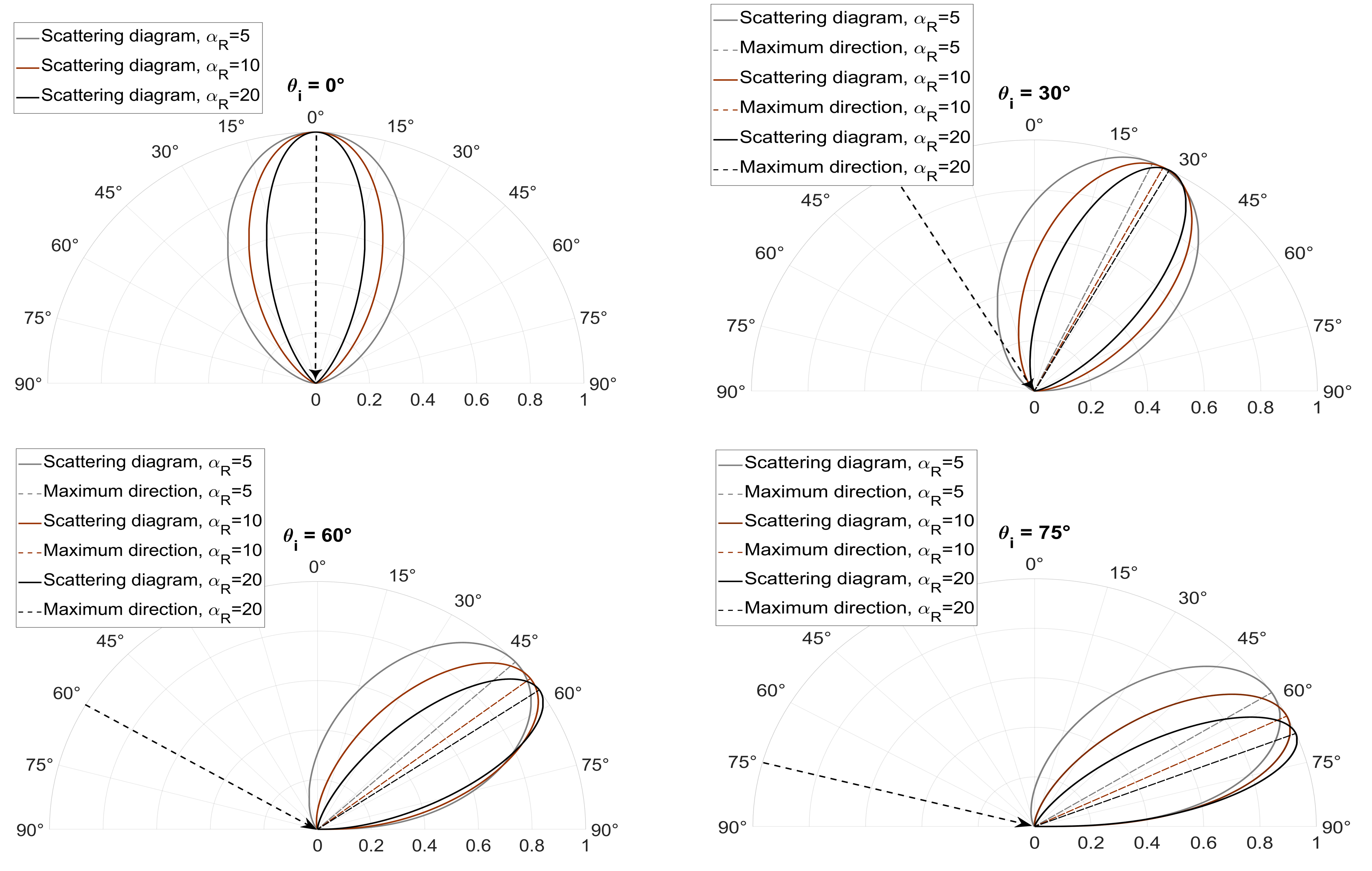}
  \caption{Scattering patterns of the new RER model, for different incidence angles and values of the parameter $\alpha_R$.}
  \label{fig:Fig_scat_diagrams}
\end{figure*}

\section{Comparisons}

The new, Reciprocal ER model (RER model in the following) is discussed and compared with the legacy ER model and other models in this section.

The shape of its scattering pattern lobe is shown in Fig. \ref{fig:Fig_scat_diagrams} for different incidence angles and $\alpha_R$ values. The lobe's directivity increases with $\alpha_R$, as it should, and its maximum is directed toward the specular direction. However, differently from the legacy ER model, the lobe is always constrained to have a null for $\theta_S=\pi/2$ in order to satisfy reciprocity as explained in section II.B. Consequently, a slight drifting of the peak away from the specular direction toward lower $\theta_S$ values can be observed for incidence angles greater than $\pi/3$ and low $\alpha_R$ values.

\begin{figure}[!ht]
  \centering
  \includegraphics[width=0.45\textwidth]{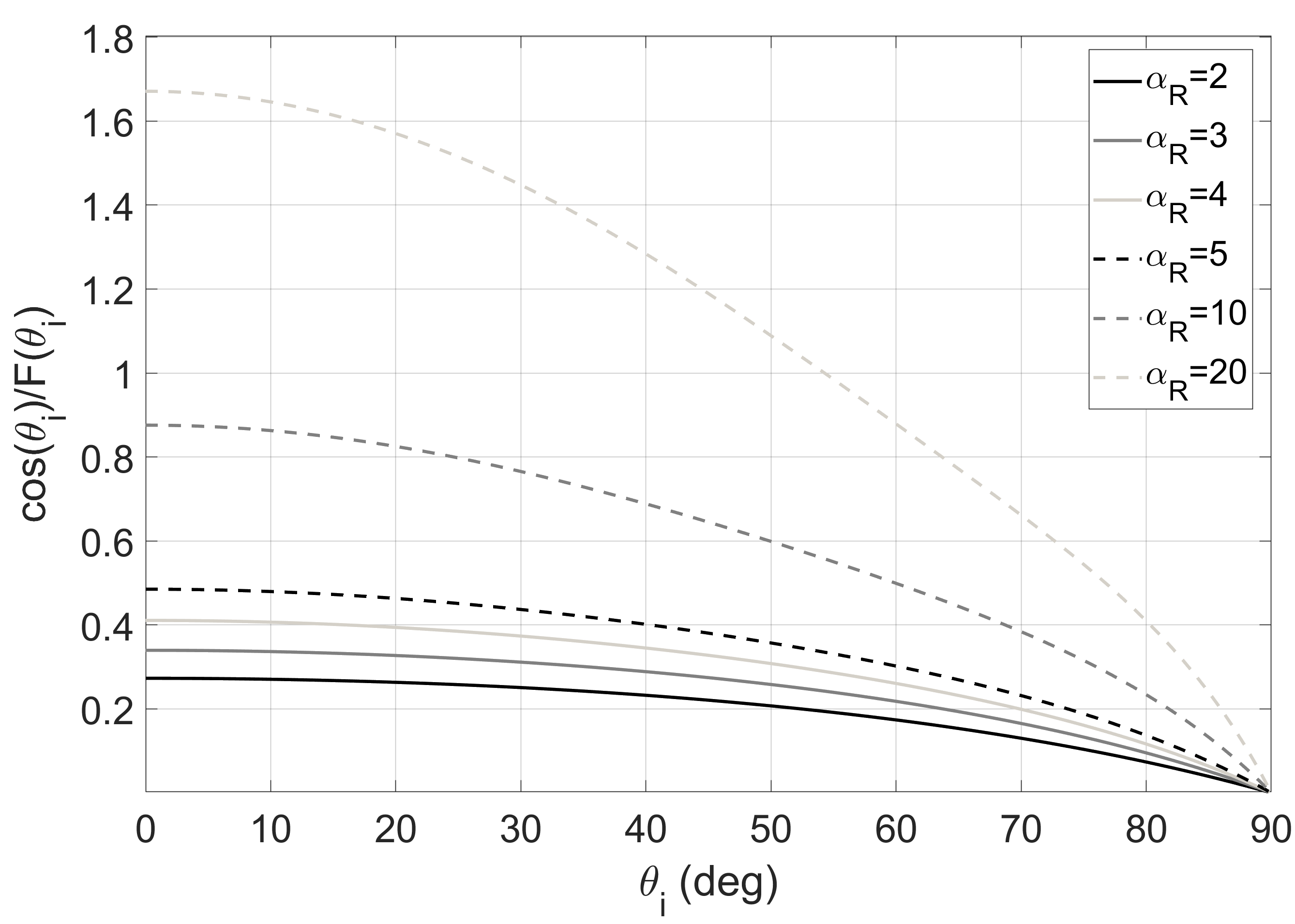}
  \caption{Evaluation of the reciprocity of the legacy ER model}
  \label{fig:legacy_model_reciprocity}
\end{figure}

As stated above, the new formulation of the ER model was derived to satisfy reciprocity. The old ER model however, was already almost reciprocal for not-too-grazing incidence angles (up to about $40^\circ$), and for low values of $\alpha_R$, as shown in Fig. \ref{fig:legacy_model_reciprocity}, where the term
$\cos\theta_i/F_{\alpha_R}\left(\theta_i,\phi_i\right)$ of eq. \eqref{eq:9} is almost constant and therefore reciprocity condition is approximately satisfied. On the other hand, it strictly respects the ER power balance, based on which it was conceived.

Vice versa, the new ER model is perfectly reciprocal but its reciprocal formulation was obtained from an approximation that slightly differs from the numerical solution of the power-balance integral \eqref{eq:8}, especially for very grazing angles of incidence, as explained in section II.B.
It is worth noting however, that reciprocity is a more important requirement than ER power-balance, since the latter is based on the simplifying yet reasonable assumption that the quantity $P_p/P_i$ of equation \eqref{eq:3} does not depend on parameter $S$, which might not be rigorously true in real-life cases.

In order to quantify the influence of the approximation in the power balance integral \eqref{eq:8}, we can introduce the \textit{power-balance anomaly}, which is normalized to the incident power $P_i$, and defined as:
\begin{equation}
\Delta_{rel}=\frac{\hat{P_s}-P_s}{P_i}
\label{eq:21}
\end{equation}
where $P_s$ is the scattered power obtained through the original power balance assumptions, i.e. through solution of the integral \eqref{eq:8} to determine the value of $F_{\alpha_R}\left(\theta_i\right)$, while $\hat{P_s}$ is the corresponding value obtained by using the approximation $F_{\alpha_R}\left(\theta_i\right) \approx k\left(\alpha_R\right)\sqrt{\cos\theta_i}$.

\begin{figure}[!ht]
  \centering
  \includegraphics[width=0.4\textwidth]{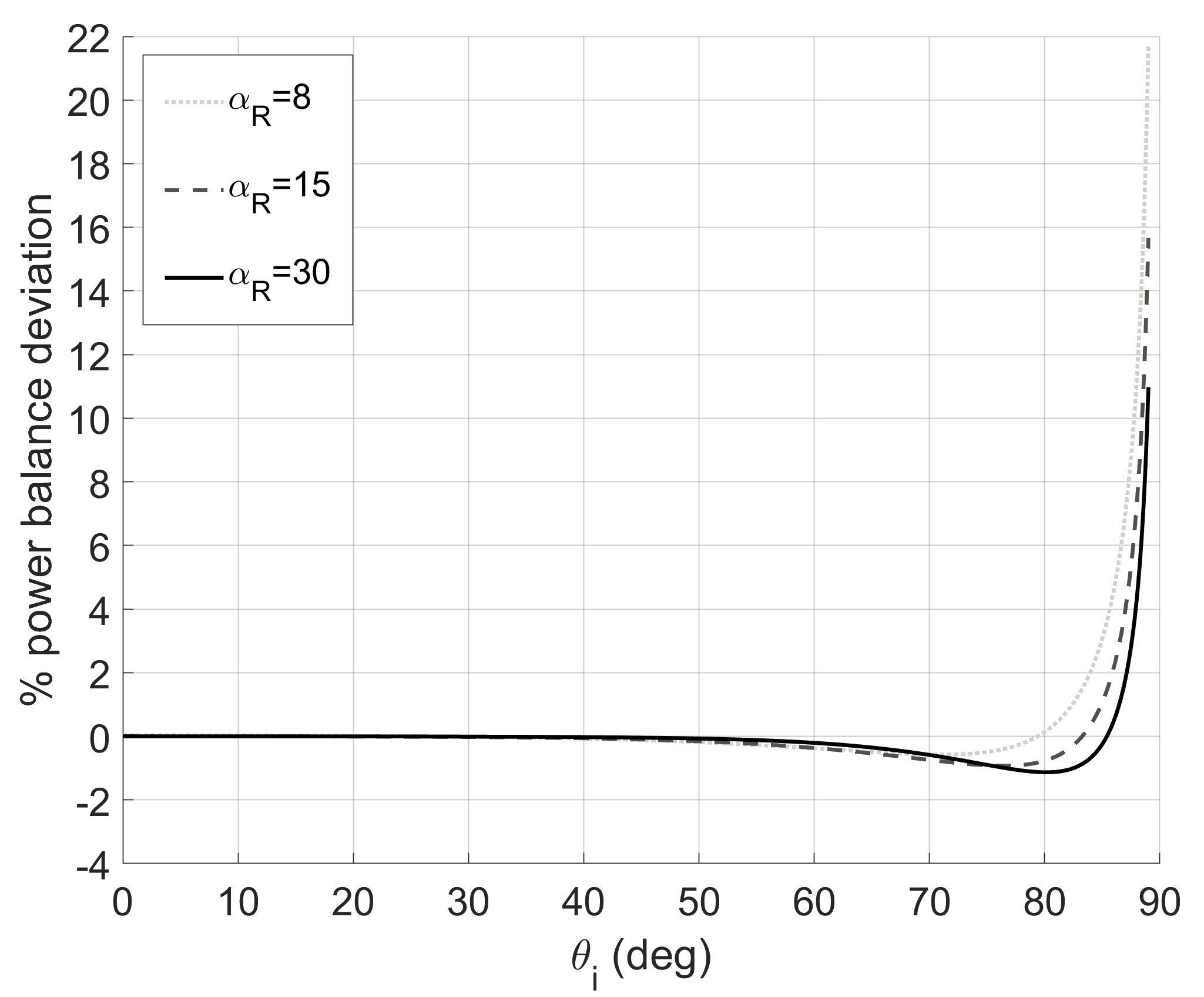}
  \caption{Percentage deviation of the reciprocal model w.r.t. legacy power balance, normalized to the incident power}
  \label{fig:power_balance_deviation}
\end{figure}

Through a few simple mathematical steps, the following expression for $\Delta_{rel}$ can be derived:
\begin{equation}
\Delta_{rel}=S^2\Gamma^2\left(\frac{F_{\alpha_R}\left(\theta_i\right)}{k\left(\alpha_R\right)\sqrt{\cos\theta_i}}-1\right)
\label{eq:22}
\end{equation}
 The power-balance anomaly $\Delta_{rel}$ of the new ER model is plotted vs. $\theta_i$ in Fig. \ref{fig:power_balance_deviation}, assuming $S=0.4$, while the modulus of the reflection coefficient $\Gamma$ was calculated with the Fresnel coefficient (TE polarization) for the case of a lossless dielectric wall with $\epsilon_r= 5$.
It is evident that $\Delta_{rel}$ is very small, within 1 \% of the incident power up to incident angles of 85 degrees or more!

It is interesting to compare the behaviour of the RER model with other models available in the literature, e.g. the Kirchhoff model for scattering from rough surfaces.
The Kirchhoff model is a widely used reference model that, being physics-based, is reciprocal and necessarily satisfies physically consistent power-balance constraints. However, it has several parameters, it is valid only for surface roughness of the Gaussian type, with correlation length larger than the wavelength, and it also has approximations that make it invalid for grazing incidence angles, as discussed for instance in \cite{Saillard2011}.

\begin{figure}[!ht]
  \centering
  \includegraphics[width=0.48\textwidth]{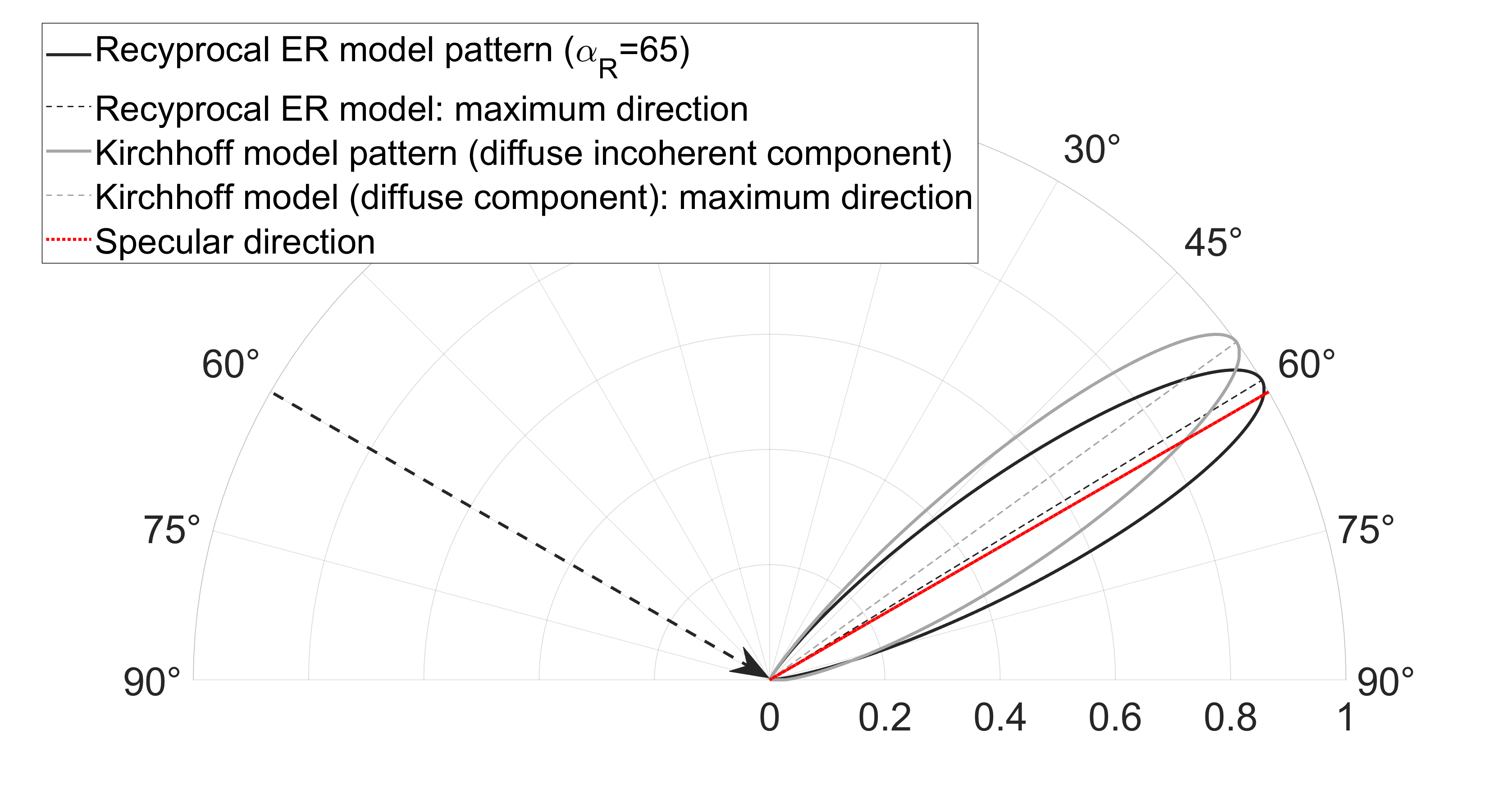}
  \caption{Comparison between Kirchhoff's DS pattern for a brick wall at 1.3 GHz and best-fit RER DS pattern ($\theta_i=45^\circ$).}
  \label{fig:Kirchhoff_comparison}
\end{figure}

The Kirchhoff's scattering coefficient provided in \cite{BeckmannSpizzichino63} is made of two parts, a coherent specular component, derived from the Radar Cross Section theory, and an incoherent diffuse component which accounts for the non-specular contribution of the facets representing the irregular surface. For the sake of comparison with the RER model, in the following we consider only the incoherent component.
In Fig. \ref{fig:Kirchhoff_comparison} the normalized scattering diagrams for the RER model and the Kirchhoff model (diffuse incoherent component) are compared for a 1.3 GHz incident plane wave with $\theta_i=\pi/3$, the same case of \cite{VDEAP2007}, Fig. 10. The following parameters are used in the Kirchhoff model: surface roughness standard deviation $\sigma_h=1\:cm$ and correlation length $l_{corr}=0.5\:m$, which are typical literature values for a brick wall as the one considered in \cite{VDEAP2007}. For comparison, the directivity parameter $\alpha_R$ of the RER model has been optimized to reproduce the same scattering lobe width as the Kirchhoff model, thus getting $\alpha_R=65$, which is a much higher value with respect to what found in \cite{VDEAP2007} for the brick wall case, i.e. $\alpha_R=4$. The shape of the two patterns is very similar: interestingly, the maximum is slightly tilted upward with respect to the specular direction in both cases, albeit to a lesser extent in the RER model case. However, the much greater degree of spreading observed in \cite{VDEAP2007} is an indication that surface elements such as indentations and material inhomogeneities (e.g. the alternation of brick and mortar, cavities inside bricks) probably give a greater contribution to DS than mere Gaussian surface roughness.
Besides the aforementioned limitations, there are additional issues that make not straightforward the implementation of the Kirchhoff model in ray-based prediction tools, as discussed for example in \cite{Priebe2011}. Moreover, as the incoherent component is computed through a series expansion, the Kirchhoff model is computational less efficient than the RER model, from 1 to 2 orders of magnitude depending on how the series is truncated.

Another possible approach to deal with DS from irregular surfaces is the one based on the use of computer graphics models, originally conceived for rendering of photorealistic images. Such models are based on the so called Bidirectional Reflectance Distribution Function (BRDF), which is a directional scattering coefficient. In recent years, "physically-based" BDRFs have been proposed, which obey to reciprocity and comply with upper bound power constraints, such as that $P_i$ should always be greater than or equal to $P_s$ \cite{Duvenhage2013}.

One example is the popular GGX shading model, originally introduced in \cite{WaMaHo2007}. In \cite{Wagen2020}, it has been proposed to use a slightly modified version of the GGX model for radio wave propagation prediction. In the model, an equivalent roughness parameter $\Sigma_s$, expressed in dB, is used: $\Sigma_s=0 \:dB$ means maximum roughness, while $\Sigma_s<-40 \:dB$ means smooth surface with quasi-specular behaviour. In particular, in \cite{Wagen2020} it is shown that, by parameterizing the GGX model to reproduce both the specular and the diffuse components and by adding them through incoherent power sum, realistic results in good agreement with the measurements can be achieved.

\begin{figure}[!ht]
  \centering
  \includegraphics[width=0.48\textwidth]{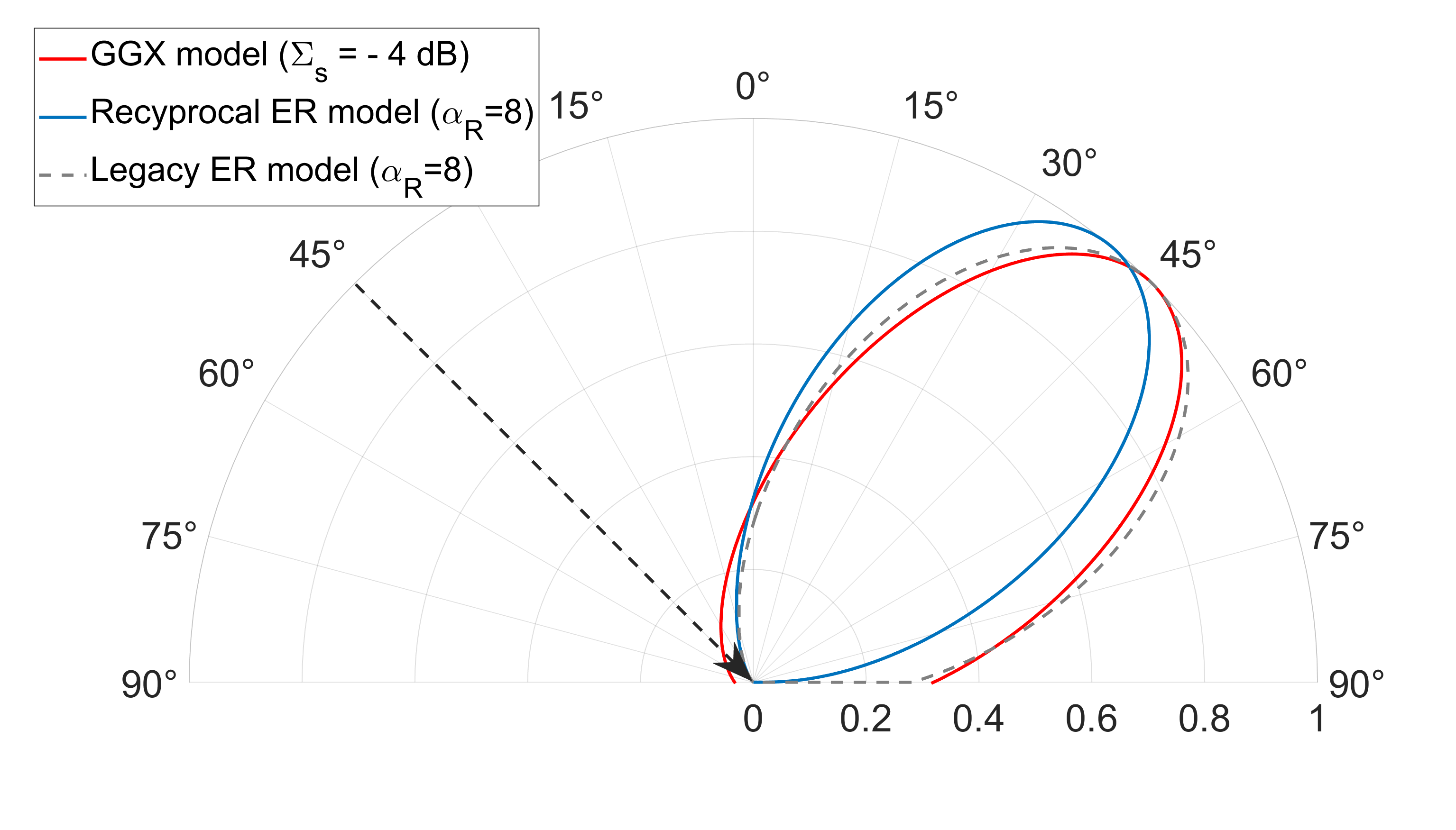}
  \caption{Comparison between GGX computer graphics model, RER model and legacy ER model for a surface with moderate roughness and $\theta_i=45^\circ$.}
  \label{fig:comparison_GGX}
\end{figure}

In Fig. \ref{fig:comparison_GGX}, the directional coefficient D of the GGX model as defined in \cite{Wagen2020}, is compared with the scattering patterns of the RER model and of the legacy ER model for an incidence angle $\theta_i=45^\circ$ and a surface with moderate roughness, i.e. $\Sigma_s=-4\:dB$. In such a case, the best-fit directivity parameter is $\alpha_R=8$ for both the RER and the legacy ER model. The GGX and the legacy ER model have a very similar scattering diagram, and interestingly both of them do not go to zero at grazing scattering angles, differently from the RER and the Kirchhoff models.

\begin{figure}[!ht]
  \centering
  \includegraphics[width=0.45\textwidth]{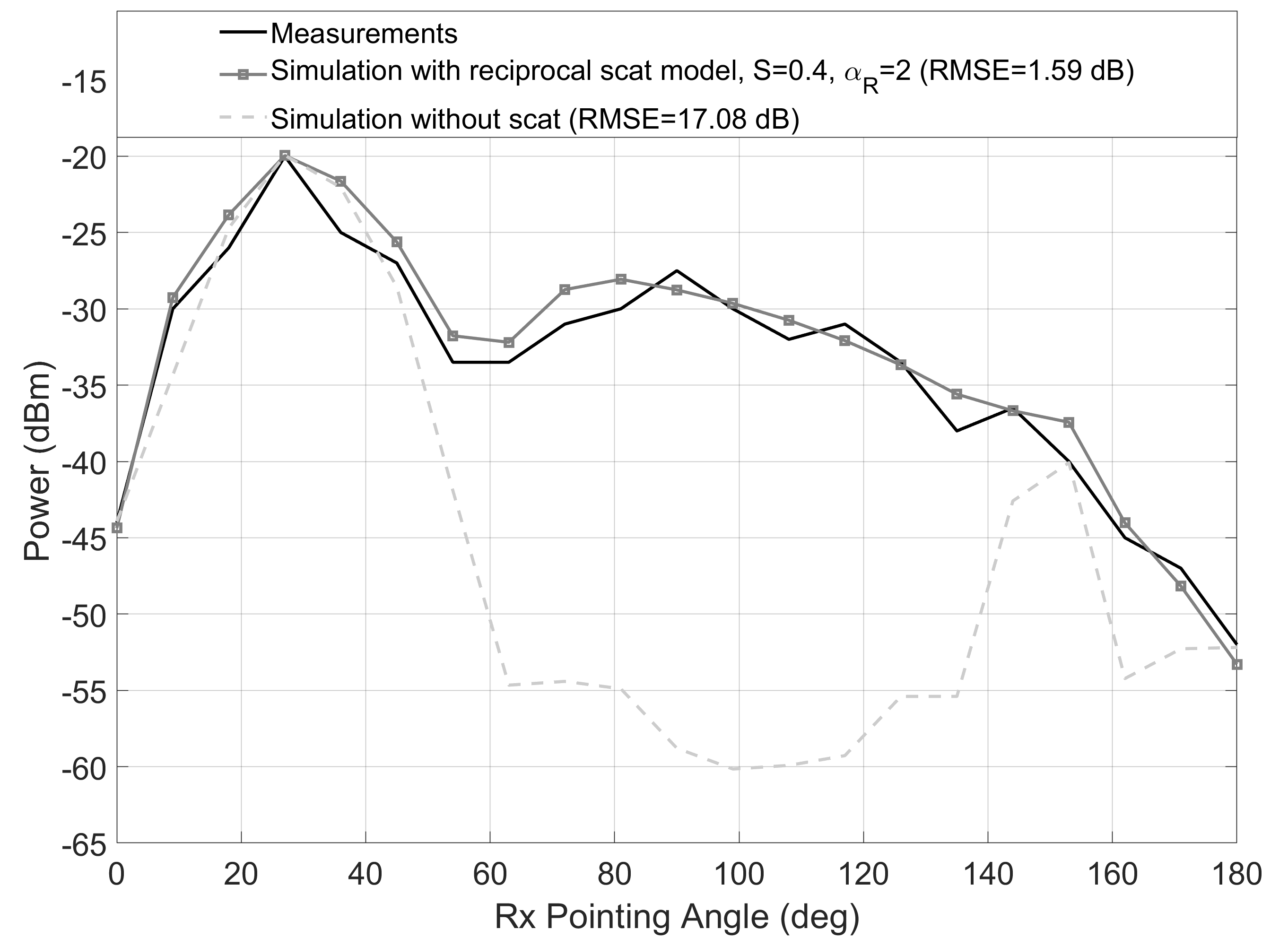}
  \caption{Comparison between RT simulation with the RER DS model embedded, and measurements in a rural building scenario, described in \cite{VDEAP2007}.}
  \label{fig:comparison_rural_building}
\end{figure}

Finally, the new ER model is compared with the measurements carried out in \cite{VDEAP2007} on a reference rural building wall. In the measurement, the façade of a rural building was illuminated by a Tx directive antenna pointing towards the centre of the wall, while the Rx directive antenna, also aiming at the wall centre, was moved along a semicircle in front of the wall to derive an estimate of the angular scattering pattern.
Despite the use of directive antennas all the interaction mechanisms (direct path, specular reflection, diffraction, diffuse scattering) are simultaneously present to some extent, and therefore the measured pattern need to be compared with RT simulations including all mechanisms: the new RER model has been embedded in the RT simulator described in \cite{VDEAP2004}, similarly to what done in \cite{VDEAP2007} for the legacy model.
Both parameters $S$ and $\alpha_R$ have been optimized to get the best match with the measured scattering pattern. Results are shown in Fig. \ref{fig:comparison_rural_building} for the optimum values of the parameters, i.e. $S=0.4$ and $\alpha_R=2$ in this case.
As expected, the scattering model allows to fill the gap for those receiving positions where the coherent interaction mechanisms (specular reflection, diffraction) are weaker. The curve corresponding to a simulation without diffuse scattering (i.e. $S=0$) is also reported for reference in the figure, and shows a very poor performance for the Rx locations further from the specular reflection angle (i.e. $\theta_i=30^\circ$ in Fig. \ref{fig:comparison_rural_building}).
From the plot, it is evident that the proposed model, if properly parametrized, can accurately describe scattering from such a typical building wall, with an RMS Error of 1.59 dB. This result is similar to, or even better than, the one shown in \cite{VDEAP2007} for the legacy ER model in the same scenario, not reported in the figure for the sake of legibility, where the best RMSE value was 1.85 dB. Similarly good results can be achieved for the other cases considered in \cite{VDEAP2007}.

\section{Conclusion}
The Effective Roughness model, a popular model for diffuse scattering from objects and building walls that is conceived to complement ray-based radio propagation models, is re-considered and modified in the present work in order to satisfy reciprocity, an important physical-soundness requisite. To this aim, the formulation of the original model has been modified to satisfy reciprocity without significantly affecting the simple and yet sound power-balance approach it is based on. The new, reciprocal version of the Effective Roughness model, which can be easily implemented and replaced to the old version in ray-based propagation models, is analyzed and compared to the old one and to other popular models in section III.
Finally, comparison with some of the measurements previously considered in \cite{VDEAP2007} for the validation of the original model has shown that the new one yields similar performance, if not better.

% or


\appendices

\section{Euler's Gamma and Beta functions}
The \textit{Euler's Gamma function} definition is \cite{abramowitz-stegun}:
\begin{equation}
\Gamma \left ( z \right )=\int_{0}^{\infty }t^{z-1}\,e^{-t}\,dt
\label{eq:A.1}
\end{equation}
where \(z\) is a complex number having positive non-zero real part (\(\Re(z)>0\)). We recap here some useful properties of the Gamma function that will be used in the proofs of the next appendices:
\begin{equation}
\Gamma \left ( z \right )= \frac{\Gamma \left (z+1  \right )}{z}
\label{eq:A.2}
\end{equation}
\begin{equation}
\Gamma \left ( z \right )\Gamma \left ( z+\frac{1}{2} \right )=2^{1-2z}\sqrt{\pi}\,\,\Gamma \left ( 2z \right )
\label{eq:A.3}
\end{equation}
\begin{equation}
\Gamma \left ( n+1 \right )=n!
\label{eq:A.4}
\end{equation}
\begin{equation}
\Gamma \left ( \frac{n}{2} \right )=\frac{(n-2)!!}{2^\frac{n-1}{2}}
\label{eq:A.5}
\end{equation}
where \(n\) is a natural number and \(n!\), \(n!!\) are the factorial and double factorial (or semi-factorial) functions, respectively, defined as:
\begin{equation}
\setlength{\jot}{6pt}
\begin{gathered}
n!=\prod_{k=0}^{n-1} \left(n-k\right)=n(n-1)(n-2)\dots\\
n!!=\prod_{k=0}^{\left \lceil \frac{n}{2} \right \rceil-1}\left ( n-2k \right )=n(n-2)(n-4)\dots
\end{gathered}
\label{eq:A.6}
\end{equation}
where $\lceil x \rceil$ stands for the least integer greater than or equal to $x$.
Also, it is conventionally assumed: $0!=1$ and $0!!=1$.
\par
In the following appendices we will also make use of the following \textit{binomial theorem} \cite{abramowitz-stegun}:
\begin{equation}
\left(a+b\right)^n=\sum\limits_{k=0}^n \binom{n}{k}\,a^{n-k}\,b^k=\sum\limits_{k=0}^n \binom{n}{k} a^k\,b^{n-k}
\label{eq:A.7}
\end{equation}
where
\begin{equation*}
\binom{n}{k}=\frac{n!}{k!\,\left(n-k\right)!}
\end{equation*}
In the particular case that $b$ is equal to the constant function $b(x)=1$, (\ref{eq:A.7}) reduces to:
\begin{equation}
\left(1+a\right)^n=\sum\limits_{k=0}^n \binom{n}{k}\,a^k
\label{eq:A.8}
\end{equation}
The \textit{Euler's Beta function} is defined as \cite{abramowitz-stegun}:
\begin{equation}
B(x,y)=\int_{0}^{1}t^{x-1}\,(1-t)^{y-1}\,dt
\label{eq:A.9}
\end{equation}
where \(\Re(x)>0\), \(\Re(y)>0\). The following properties hold:
\begin{equation}
B(x,y)=\frac{\Gamma \left ( x \right ) \Gamma \left ( y \right )}{\Gamma \left ( x+y \right ) }
\label{eq:A.10}
\end{equation}
\begin{equation}
B(x,y)=2\int_{0}^{\pi/2}\sin^{2x-1}t \,\, \cos^{2y-1}t\,\,dt
\label{eq:A.11}
\vspace{5mm}
\end{equation}

\section{
Complete Derivation of $F_{\alpha_R}$ for the legacy ER model (non-reciprocal)}

The aim of this section is to prove equation (\ref{eq:13}), which is the closed-form solution of the integral (\ref{eq:8}) for the single-lobe scattering pattern of the legacy ER model (see eq. (\ref{eq:10})), originally proposed in \cite{VDEAP2007}.
The integral to be solved is:
\begin{equation}
F_{\alpha_R}=\int_{0}^{2\pi}\int_{0}^{\pi/2}\left ( \frac{1+\cos{\psi_R}}{2} \right )^{\alpha_R}\sin \theta_S\,d\theta_S\,d\phi_S
\label{eq:B.1}
\end{equation}
with \(\cos\psi_R=\cos\theta_i\cos\theta_S-\sin\theta_i\sin\theta_S\cos\left ( \phi_S-\phi_i \right )\). Using the binomial theorem in the form (\ref{eq:A.7}) we obtain
\begin{equation}
F_{\alpha_R}=\frac{1}{2^{\alpha_R}} \int_{0}^{2\pi} \int_{0}^{\pi/2}\sum_{j=0}^{\alpha_R}\binom{\alpha_R}{j}\cos^{j}\psi_R\sin\theta_S\,d\theta_S\,d\phi_S
\label{eq:B.2}
\end{equation}
It follows that
\begin{equation}
\begin{split}
F_{\alpha_R}=\frac{1}{2^{\alpha_R}}\sum_{j=0}^{\alpha_R}\binom{\alpha_R}{j} \int_{0}^{2\pi}\int_{0}^{\pi/2}( \cos\theta_i\cos\theta_S+\\-\sin\theta_i\sin\theta_S\cos\left ( \phi_S-\phi_i \right )  )^j \sin\theta_S\,d\theta_S\,d\phi_S
\label{eq:B.3}
\end{split}
\end{equation}
By applying the binomial theorem again (eq.(\ref{eq:A.8})), we get
\begin{equation}
\begin{split}
F_{\alpha_R}=\frac{1}{2^{\alpha_R}}\sum_{j=0}^{\alpha_R}\binom{\alpha_R}{j} \sum_{l=0}^{j}\binom{j}{l}(-1)^l \cos^{j-l}\theta_i\sin^{l}\theta_i \: \cdot \\ \cdot\int_{0}^{2\pi}\int_{0}^{\pi/2}\cos^{j-l}\theta_S\sin^{l}\theta_S\cos^{l}\left ( \phi_S-\phi_i \right )\sin\theta_S\,d\theta_S\,d\phi_S
\label{eq:B.4}
\end{split}
\end{equation}
Let \(I\) stand for
\begin{equation}
\begin{split}
I&=\int_{0}^{2\pi}\int_{0}^{\pi/2}\cos^{j-l}\theta_S\sin^{l}\theta_S\cos^{l}\left ( \phi_S-\phi_i \right )\sin\theta_S\,d\theta_S\,d\phi_S\\&=\int_{0}^{\pi/2}\cos^{j-l}\theta_S\sin^{l+1}\theta_S\,d\theta_S\, \int_{0}^{2\pi} \cos^{l}\left ( \phi_S-\phi_i \right )\,d\phi_S\\&=I_1\cdot I_2
\label{eq:B.5}
\end{split}
\end{equation}
Let us consider the first factor in (\ref{eq:B.5}), i.e. \(I_1=\int_{0}^{\pi/2}\cos^{j-l}\theta_S\sin^{l+1}\theta_S\,d\theta_S\).
Applying (\ref{eq:A.11}) we obtain:
\begin{equation}
I_1=\frac{1}{2}B\left ( \frac{l}{2}+1,\frac{j-l+1}{2} \right )
\label{eq:B.6}
\end{equation}
The second factor \(I_2=\int_{0}^{2\pi} \cos^{l}\left ( \phi_S-\phi_i \right )\,d\phi_S\) may be written as
\begin{equation}
I_2=\int_{0}^{2\pi} \left ( \cos\phi_S  \cos\phi_i +  \sin\phi_S \sin\phi_i\right )^l\,d\phi_S
\label{eq:B.7}
\end{equation}
Using the binomial theorem we get:
\begin{equation}
I_2=\sum_{q=0}^{l}\binom{l}{q}\cos^{l-q}\phi_i \sin^q\phi_i \int_{0}^{2\pi} \cos^{l-q}\phi_S \sin^q\phi_S \,d\phi_S
\label{eq:B.8}
\end{equation}
Let be \(X=\int_{0}^{2\pi} \cos^{l-q}\phi_S \sin^q\phi_S \,d\phi_S\). This integral can be split into four parts:
\begin{equation}
\begin{split}
X=&\int_{0}^{\pi/2} \cos^{l-q}\phi_S \sin^q\phi_S \,d\phi_S\,+\\&+\int_{\pi/2}^{\pi} \cos^{l-q}\phi_S \sin^q\phi_S \,d\phi_S\,+ \\ &+\int_{\pi}^{3\pi/2} \cos^{l-q}\phi_S \sin^q\phi_S \,d\phi_S\,+\\&+\int_{3\pi/2}^{2\pi} \cos^{l-q}\phi_S \sin^q\phi_S \,d\phi_S
\label{eq:B.9}
\end{split}
\end{equation}
It is evident that if \(q\) is odd the four terms cancel each other out. The same thing happens when \(q\) is even and \(l\) is odd. On the other hand when \(l\) and \(q\) are both even, the four terms give the same value. Therefore we have
\begin{equation}
X=\left\{\begin{matrix}
4\int_{0}^{\pi/2} \cos^{l-q}\phi_S \sin^q\phi_S \,d\phi_S, & \textrm{if \textit{l,q} even}\\
0 & \textrm{otherwise}
\end{matrix}\right.
\label{eq:B.10}
\end{equation}
Using (\ref{eq:A.11}) we obtain:
\begin{equation}
X=\left\{\begin{matrix}
2B\left ( \frac{q+1}{2},\frac{l-q+1}{2} \right ), & \textrm{if \textit{l,q} even}\\
0 & \textrm{otherwise}
\end{matrix}\right.
\label{eq:B.11}
\end{equation}
Combining (\ref{eq:B.4}), (\ref{eq:B.5}), (\ref{eq:B.6}), (\ref{eq:B.8}) we can assert that
\begin{equation}
\begin{split}
F_{\alpha_R}=&\frac{1}{2^{\alpha_R}}\sum_{j=0}^{\alpha_R}\binom{\alpha_R}{j} \sum_{l=0}^{j}\binom{j}{l}(-1)^l \cos^{j-l}\theta_i\sin^{l}\theta_i\, \cdot \\ &\cdot \frac{1}{2}B\left ( \frac{l}{2}+1,\frac{j-l+1}{2} \right )\sum_{q=0}^{l}\binom{l}{q}\cos^{l-q}\phi_i \sin^q\phi_i \,X
\end{split}
\label{eq:B.12}
\end{equation}
Since \(X\) is nonzero only when the indices \(l\) and \(q\) are even, combining (\ref{eq:B.11}) and (\ref{eq:B.12}) we can write
\begin{equation}
\begin{split}
F_{\alpha_R}=\frac{1}{2^{\alpha_R}}\sum_{j=0}^{\alpha_R}\binom{\alpha_R}{j} \sum_{l=0}^{\left \lfloor j/2 \right \rfloor}\binom{j}{2l} \cos^{j-2l}\theta_i\sin^{2l}\theta_i\, \cdot \\ \cdot B\left ( l+1,\frac{j-2l+1}{2} \right ) \sum_{q=0}^{l}\binom{2l}{2q}\cos^{2l-2q}\phi_i \sin^{2q}\,\phi_i \, \cdot \\ \cdot B\left ( q+\frac{1}{2},l-q+\frac{1}{2} \right )
\end{split}
\label{eq:B.13}
\end{equation}
Let us consider \(B\left ( l+1,\frac{j-2l+1}{2} \right )\) and \(B\left ( q+\frac{1}{2},l-q+\frac{1}{2} \right )\). Using properties (\ref{eq:A.10}), (\ref{eq:A.5}), (\ref{eq:A.4}) it is simple to obtain
\begin{equation}
B\left ( l+1,\frac{j-2l+1}{2} \right )=2^{l+1}\,l!\,\frac{\left ( j-2l-1 \right )!!}{\left ( j+1 \right )!!}
\label{eq:B.14}
\end{equation}
In a similar way, using (\ref{eq:A.10}), (\ref{eq:A.4}), (\ref{eq:A.3}) we get
\begin{equation}
B\left ( q+\frac{1}{2},l-q+\frac{1}{2} \right )=\frac{\pi\left ( 2q \right )!\left ( 2l-2q \right )!}{2^{2l}\,q!\,l!\left ( l-q \right )!}
\label{eq:B.15}
\end{equation}
By substituting (\ref{eq:B.14}) and (\ref{eq:B.15}) into (\ref{eq:B.13}) and rearranging some terms we obtain
\begin{equation}
\begin{split}
F_{\alpha_R}=&\frac{2\pi\,\alpha_R!}{2^{\alpha_R}}\sum_{j=0}^{\alpha_R}\frac{1}{\left ( \alpha_R-j \right )!\left ( j+1 \right )!!}\sum_{l=0}^{\left \lfloor j/2 \right \rfloor}\frac{\cos^{j-2l}\theta_i\sin^{2l}\theta_i}{2^{l}\left ( j-2l \right )!!} \cdot \\& \cdot \sum_{q=0}^{l}\frac{\cos^{2l-2q}\phi_i \sin^{2q}\phi_i}{q!\left ( l-q \right )!}
\label{eq:B.16}
\end{split}
\end{equation}
where the symbol $\lfloor x \rfloor$ stands for the greatest integer less than or equal to $x$.

But \(\sum_{q=0}^{l}\frac{\cos^{2l-2q}\phi_i \sin^{2q}\phi_i}{q!\left ( l-q \right )!}\) equals \(1/l!\), as it can be easily verified by applying the binomial theorem. Then, after this substitution we can eventually write the final formulation of \(F_{\alpha_R}\):
\begin{equation}
F_{\alpha_R}=\frac{2\pi\,\alpha_R!}{2^{\alpha_R}}\sum_{j=0}^{\alpha_R}\frac{1}{\left ( \alpha_R-j \right )!\left ( j+1 \right )!!}\sum_{l=0}^{\left \lfloor j/2 \right \rfloor}\frac{\cos^{j-2l}\theta_i\sin^{2l}\theta_i}{2^{l}\,l!\left ( j-2l \right )!!}
\label{eq:B.17}
\end{equation}

This equation is also valid in the case of normal incidence, i.e. $\theta_i=0$, if for the $0^{th}$-order term of the second summation it is \textit{conventionally} assumed, as most automatic calculators do:
\begin{equation*}
\lim_{\theta_i \to 0}\left(\sin\theta_i\right)^0=0^0=1
\end{equation*}
In such a case, for normal incidence (\ref{eq:B.17}) reduces to:
\begin{equation*}
F_{\alpha_R}\left(0\right)=\frac{4\pi}{\alpha_R+1}\left(1-\frac{1}{2^{\alpha_R+1}}\right)
\end{equation*}
The same result can be also obtained by directly integrating (\ref{eq:8}) for $\theta_i=0$, which is straightforward.

\section{Solution of the power balance integral for the new ER reciprocal model (single-lobe version)}

The aim of this section is to justify equation (\ref{eq:17}).
We assume that $F_{\alpha_R}$ can be written in the following, approximate form, as discussed in Section II:
\begin{equation}
F_{\alpha_R} \approx k\left ( \alpha_R \right )\sqrt{\cos\theta_i}
\label{eq:C.1}
\end{equation}
Looking at this expression, we observe that \(k\left ( \alpha_R \right )\) is equal to \(F_{\alpha_R}\) when \(\theta_i=0\). Thus we can assert that
\begin{equation}
F_{\alpha_R} \approx k\left ( \alpha_R \right )\sqrt{\cos\theta_i}=F_{\alpha_R}\left ( \theta_i=0 \right )\sqrt{\cos\theta_i}
\label{eq:C.2}
\end{equation}
The value of $F_{\alpha_R}\left ( \theta_i=0 \right )$ can be found by directly integrating (\ref{eq:8}) for $\theta_i=0$, and observing that, for normal incidence, $\cos \psi_R=\cos \theta_S$. Then, we get:
\begin{equation}
F_{\alpha_R} \approx \frac{2\pi\sqrt{\cos\theta_i}}{2^{\alpha_R}}\int_{0}^{\frac{\pi}{2}} \sqrt{\cos\theta_S}\left (1+\cos\theta_S\right )^{\alpha_R}\sin\theta_S\:d\theta_S
\label{eq:C.3}
\end{equation}
Then, by applying the binomial theorem we obtain
\begin{equation}
F_{\alpha_R} \approx \frac{2\pi\sqrt{\cos\theta_i}}{2^{\alpha_R}}\int_{0}^{\frac{\pi}{2}} \sin\theta_S\sqrt{\cos\theta_S}\sum_{j=0}^{\alpha_R}\binom{\alpha_R}{j}\cos^{j}\theta_S\:d\theta_S
\label{eq:C.4}
\end{equation}
Equation (\ref{eq:C.4}) can be rewritten as
\begin{equation}
F_{\alpha_R} \approx \frac{2\pi\sqrt{\cos\theta_i}}{2^{\alpha_R}}\sum_{j=0}^{\alpha_R}\binom{\alpha_R}{j}\int_{0}^{\frac{\pi}{2}} \sin\theta_S\cos^{j+\frac{1}{2}}\theta_S\:d\theta_S
\label{eq:C.5}
\end{equation}
Applying (\ref{eq:A.11}) in (\ref{eq:C.5}) we have
\begin{equation}
F_{\alpha_R} \approx \frac{\pi\sqrt{\cos\theta_i}}{2^{\alpha_R}}\sum_{j=0}^{\alpha_R}\binom{\alpha_R}{j}B\left ( 1,\frac{j}{2}+\frac{3}{4} \right )
\label{eq:C.6}
\end{equation}
Using properties (\ref{eq:A.10}), (\ref{eq:A.2}) we get the final expression for \(F_{\alpha_R}\):
\begin{equation}
F_{\alpha_R} \approx \frac{4\pi\sqrt{\cos\theta_i}}{2^{\alpha_R}}\sum_{j=0}^{\alpha_R}\binom{\alpha_R}{j}\frac{1}{2j+3}
\label{eq:C.7}
\end{equation}
Eventually, eq. (\ref{eq:17}) is obtained by substituting (\ref{eq:C.7}) and (\ref{eq:14}) into (\ref{eq:7}).

\section{Solution of the power balance integral for the double-lobe RER model}

The aim of this section is to justify equation (\ref{eq:20}). In order to do that, let's consider the two integrals in (\ref{eq:19})
\begin{equation}
\setlength{\jot}{10pt}
\begin{gathered}
F_{\alpha_R}=\int\limits_{0}^{2\pi}\int\limits_{0}^{\frac{\pi}{2}} \sqrt{\cos\theta_S} \, \left (\frac{1+\cos\psi_R}{2}\right )^{\alpha_R}     \sin\theta_S d \theta_S d \phi_S\\
F_{\alpha_i}=\int\limits_{0}^{2\pi}\int\limits_{0}^{\frac{\pi}{2}} \sqrt{\cos\theta_S} \, \left (\frac{1+\cos\psi_i}{2}\right )^{\alpha_i} \sin\theta_S d\theta_S d \phi_S
\end{gathered}
\label{eq:D.1}
\end{equation}
and remember that \cite{VDEAP2007}:
\begin{equation}
\cos\psi_i=\cos\theta_i\cos\theta_S+\sin\theta_i\sin\theta_S\cos\left ( \phi_S-\phi_i \right )
\label{eq:D.2}
\end{equation}
We notice that only difference between (\ref{eq:11}) and (\ref{eq:D.2}) is in the sign of the second term.

We want to show that the integrals above have the same result, for a fixed value of the exponent.
Proving this is equivalent to proving that the result of the integral \(M\) defined in equation (\ref{eq:D.3}) does not depend on the sign \(\pm\) of the term \(\sin\theta_i\sin\theta_S\cos\left ( \phi_S-\phi_i \right ) \):
\begin{equation}
\begin{split}
M=&\int_{0}^{2\pi}\int_{0}^{\pi/2} \sqrt{\cos\theta_S}(1+\cos\theta_i \cos\theta_S+ \\ &\pm \sin\theta_i\sin\theta_S\cos\left ( \phi_S-\phi_i \right ))^{m}\sin \theta_S\,d\theta_{S}\,d\phi_{S}
\end{split}
\label{eq:D.3}
\end{equation}
In order to prove that, we apply the binomial theorem twice and with a little manipulation of the factors we get
\begin{equation}
\begin{split}
M=\sum_{j=0}^{m}\binom{m}{j}\sum_{l=0}^{j}[\binom{j}{l}\left ( \pm 1 \right )^l \sin^l \theta_i \cos^{j-l}\theta_i\,\cdot \\ \cdot\int_{0}^{\pi/2}\sin^{l+1}\theta_S \cos^{j-l+\frac{1}{2}}\theta_S \,d\theta_{S}\,\cdot \\ \cdot\int_{0}^{2\pi}\cos^l\left ( \phi_S-\phi_i \right )^{\alpha_R}\sin \theta_S\,d\phi_{S}]
\end{split}
\label{eq:D.4}
\end{equation}
But \(\int_{0}^{2\pi}\cos^l\left ( \phi_S-\phi_i \right )^{\alpha_R}\sin \theta_S\,d\phi_{S}\neq 0\) only when the index \(l\) is even. Thus, the overall contribution of the term \(\left (\pm1  \right )^l\) is completely irrelevant since \(\left (+1  \right )^l=\left (-1  \right )^l=1\) when \(l\) is even. This proves that the sign \(\pm\) does not change the result of \(M\).

Therefore, the 2 integrals in (\ref{eq:D.1}) have the same form, and the only difference between them is in the value of the exponent, either $\alpha_R$ or $\alpha_i$.
By adopting the same procedure as in Appendix C, we then find that:
\begin{equation}
\begin{gathered}
F_{\alpha_R} \approx F_{\alpha_R}\left (0 \right )\sqrt{\cos\theta_i} \approx \frac{4\pi\sqrt{\cos\theta_i}}{2^{\alpha_R}}\sum_{j=0}^{\alpha_R}\binom{\alpha_R}{j}\frac{1}{2j+3}\\
F_{\alpha_i} \approx F_{\alpha_i}\left (0 \right )\sqrt{\cos\theta_i} \approx \frac{4\pi\sqrt{\cos\theta_i}}{2^{\alpha_i}}\sum_{j=0}^{\alpha_i}\binom{\alpha_i}{j}\frac{1}{2j+3}
\end{gathered}
\label{eq:D.5}
\end{equation}
and by substituting $F_{\alpha_i,\alpha_R}=\Lambda F_{\alpha_R}+(1-\Lambda)F_{\alpha_i}$ into (\ref{eq:18}), we finally get (\ref{eq:20}).

%\section*{Acknowledgment}


\ifCLASSOPTIONcaptionsoff
  \newpage
\fi



\bibliographystyle{IEEEtran}

\bibliography{bibtex/bib/IEEEabrv,bibtex/bib/references}

\begin{thebibliography}{10}
\providecommand{\url}[1]{#1}
\csname url@samestyle\endcsname
\providecommand{\newblock}{\relax}
\providecommand{\bibinfo}[2]{#2}
\providecommand{\BIBentrySTDinterwordspacing}{\spaceskip=0pt\relax}
\providecommand{\BIBentryALTinterwordstretchfactor}{4}
\providecommand{\BIBentryALTinterwordspacing}{\spaceskip=\fontdimen2\font plus
\BIBentryALTinterwordstretchfactor\fontdimen3\font minus
  \fontdimen4\font\relax}
\providecommand{\BIBforeignlanguage}[2]{{%
\expandafter\ifx\csname l@#1\endcsname\relax
\typeout{** WARNING: IEEEtran.bst: No hyphenation pattern has been}%
\typeout{** loaded for the language `#1'. Using the pattern for}%
\typeout{** the default language instead.}%
\else
\language=\csname l@#1\endcsname
\fi
#2}}
\providecommand{\BIBdecl}{\relax}
\BIBdecl

\bibitem{Felsen1973}
L.~Felsen and N.~Marcuvitz, \emph{Radiation and Scattering of Waves}.\hskip 1em
  plus 0.5em minus 0.4em\relax New York: IEEE Press, 1973.

\bibitem{DEB1999}
V.~Degli-Esposti and H.~Bertoni, ``Evaluation of the role of diffuse scattering
  in urban microcellular propagation,'' in \emph{Gateway to 21st Century
  Communications Village. VTC 1999-Fall. IEEE VTS 50th Vehicular Technology
  Conference (Cat. No.99CH36324)}, vol.~3, 1999, pp. 1392--1396 vol.3.

\bibitem{VDEAP2001}
V.~Degli-Esposti, ``A diffuse scattering model for urban propagation
  prediction,'' \emph{{IEEE} Trans. Antennas Propag.}, vol.~49, no.~7, pp.
  1111--1113, 2001.

\bibitem{VDEAP2007}
V.~Degli-Esposti, F.~Fuschini, E.~M. Vitucci, and G.~Falciasecca, ``Measurement
  and modelling of scattering from buildings,'' \emph{{IEEE} Trans. Antennas
  Propag.}, vol.~55, no.~1, pp. 143--153, 2007.

\bibitem{vitucci2018scalemodel}
E.~M. Vitucci, J.~Chen, V.~Degli-Esposti, J.~S. Lu, H.~L. Bertoni, and X.~Yin,
  ``Analyzing radio scattering caused by various building elements using
  millimeter-wave scale model measurements and ray tracing,'' \emph{{IEEE}
  Trans. Antennas Propag.}, vol.~67, no.~1, pp. 665--669, 2018.

\bibitem{VDEAP2004}
V.~Degli-Esposti, D.~Guiducci, A.~de'Marsi, P.~Azzi, and F.~Fuschini, ``An
  advanced field prediction model including diffuse scattering,'' \emph{{IEEE}
  Trans. Antennas Propag.}, vol.~52, no.~7, pp. 1717--1728, 2004.

\bibitem{Mani2012}
F.~Mani, F.~Quitin, and C.~Oestges, ``Directional spreads of dense multipath
  components in indoor environments: Experimental validation of a ray-tracing
  approach,'' \emph{{IEEE} Trans. Antennas Propag.}, vol.~60, no.~7, pp.
  3389--3396, 2012.

\bibitem{vitucci2012polarimetric}
E.~M. Vitucci, F.~Mani, V.~Degli-Esposti, and C.~Oestges, ``Polarimetric
  properties of diffuse scattering from building walls: Experimental
  parameterization of a ray-tracing model,'' \emph{{IEEE} Trans. Antennas
  Propag.}, vol.~60, no.~6, pp. 2961--2969, 2012.

\bibitem{mani2014parameterization}
F.~Mani, E.~M. Vitucci, F.~Quitin, V.~Degli-Esposti, and C.~Oestges,
  ``Parameterization of a polarimetric diffuse scattering model in indoor
  environments,'' \emph{{IEEE} Trans. Antennas Propag.}, vol.~62, no.~8, pp.
  4361--4364, 2014.

\bibitem{vitucci2014MIMO}
E.~M. Vitucci, L.~Tarlazzi, F.~Fuschini, P.~Faccin, and V.~Degli-Esposti,
  ``Interleaved-{MIMO} {DAS} for indoor radio coverage: Concept and performance
  assessment,'' \emph{{IEEE} Trans. Antennas Propag.}, vol.~62, no.~6, pp.
  3299--3309, 2014.

\bibitem{VDE2014RTbeamforming}
V.~Degli-Esposti, F.~Fuschini, E.~M. Vitucci, M.~Barbiroli, M.~Zoli, L.~Tian,
  X.~Yin, D.~Dupleich, R.~Muller, C.~Schneider \emph{et~al.}, ``Ray-tracing
  based mm-wave beamforming assessment,'' \emph{IEEE Access}, vol.~2, 2014.

\bibitem{fuschini2019studybeamforming}
F.~Fuschini, M.~Zoli, E.~M. Vitucci, M.~Barbiroli, and V.~Degli-Esposti, ``A
  study on millimeter-wave multiuser directional beamforming based on
  measurements and ray tracing simulations,'' \emph{{IEEE} Trans. Antennas
  Propag.}, vol.~67, no.~4, pp. 2633--2644, 2019.

\bibitem{Tengfei2019}
E.~M. Vitucci, V.~Degli-Esposti, F.~Mani, F.~Fuschini, M.~Barbiroli, M.~Gan,
  C.~Li, J.~Zhao, and Z.~Zhong, ``Tuning ray tracing for mm-wave coverage
  prediction in outdoor urban scenarios,'' \emph{Radio Science}, vol.~54,
  no.~11, pp. 1112--1128, 2019.

\bibitem{SheikhMIMOTHz}
F.~Sheikh, Y.~Gao, and T.~Kaiser, ``A study of diffuse scattering in massive
  {MIMO} channels at terahertz frequencies,'' \emph{{IEEE} Trans. Antennas
  Propag.}, vol.~68, no.~2, pp. 997--1008, 2020.

\bibitem{XieDiffuseTHz2022}
P.~Xie, K.~Guan, D.~He, H.~Yi, J.~Dou, and Z.~Zhong, ``Terahertz wave
  propagation characteristics on rough surfaces based on full-wave
  simulations,'' \emph{Radio Science}, vol.~57, no.~6, 2022.

\bibitem{JansenTHz2011}
C.~Jansen, S.~Priebe, C.~M\"oller, M.~Jacob, H.~Dierke, M.~Koch, and
  T.~K\"urner, ``Diffuse scattering from rough surfaces in {THz} communication
  channels,'' \emph{IEEE Trans. Terahertz Sci. Technol.}, vol.~1, no.~2, pp.
  462--472, 2011.

\bibitem{BeckmannSpizzichino63}
P.~Beckmann and A.~Spizzichino, \emph{The scattering of electromagnetic waves
  from rough surfaces}.\hskip 1em plus 0.5em minus 0.4em\relax Oxford: Pergamon
  Press, 1963.

\bibitem{Tsang2000}
L.~Tsang, J.~Kong, and K.-H. Ding, \emph{Scattering of Electromagnetic Waves:
  Theories and Applications}.\hskip 1em plus 0.5em minus 0.4em\relax Hoboken,
  New Jersey: John Wiley and Sons, Ltd, 2000.

\bibitem{Bertoni2004}
P.~Pongsilamanee and H.~L. Bertoni, ``Specular and nonspecular scattering from
  building fa\c{c}ades,'' \emph{{IEEE} Trans. Antennas Propag.}, vol.~52,
  no.~7, pp. 1879--1889, 2004.

\bibitem{DeFuVi2009}
V.~Degli-Esposti, F.~Fuschini, and E.~M. Vitucci, ``A fast model for
  distributed scattering from buildings,'' in \emph{2009 3rd European
  Conference on Antennas and Propagation}, 2009, pp. 1932--1936.

\bibitem{FuDeVi2010}
F.~Fuschini, V.~Degli-Esposti, and E.~M. Vitucci, ``A model for forward-diffuse
  scattering through a wall,'' in \emph{Proceedings of the Fourth European
  Conference on Antennas and Propagation}, 2010, pp. 1--4.

\bibitem{MiDeDe2014}
L.~Minghini, R.~D'Errico, V.~Degli~Esposti, and E.~M. Vitucci,
  ``Electromagnetic simulation and measurement of diffuse scattering from
  building walls,'' in \emph{The 8th European Conference on Antennas and
  Propagation (EuCAP 2014)}, 2014, pp. 1298--1302.

\bibitem{Pascual2016}
J.~Pascual-Garc\'ia, J.-M. Molina-Garc\'ia-Pardo, M.-T. Mart\'inez-Ingl\'es,
  J.-V. Rodr\'iguez, and N.~Saur\'in-Serrano, ``On the importance of diffuse
  scattering model parameterization in indoor wireless channels at mm-wave
  frequencies,'' \emph{IEEE Access}, vol.~4, pp. 688--701, 2016.

\bibitem{ItemLevel2016}
F.~Fuschini, S.~H\"afner, M.~Zoli, R.~M\"uller, E.~M. Vitucci, D.~Dupleich,
  M.~Barbiroli, J.~Luo, E.~Schulz, V.~Degli-Esposti, and R.~S. Thom\"a, ``Item
  level characterization of mm-wave indoor propagation,'' \emph{Eurasip J.
  Wirel. Commun. Netw.}, vol. 2016, no.~4, 2016.

\bibitem{Remcom}
Remcom. (2022) Wireless {InSite} {Diffuse} {Scattering}. Available:\\
  \url{https://www.remcom.com/wireless-insite-diffuse-scattering}. Accessed:
  2022-08-09.

\bibitem{VanBladel07}
J.~Van~Bladel, \emph{Electromagnetic Fields}.\hskip 1em plus 0.5em minus
  0.4em\relax New York: Wiley-IEEE Press, 2007, pp. 405--413.

\bibitem{Duvenhage2013}
B.~Duvenhage, K.~Bouatouch, and D.~G. Kourie, ``Numerical verification of
  bidirectional reflectance distribution functions for physical plausibility,''
  in \emph{Proceedings of the South African Institute for Computer Scientists
  and Information Technologists Conference}, ser. SAICSIT '13, 2013, p.
  200–208.

\bibitem{WaMaHo2007}
B.~Walter, S.~Marschner, H.~Li, and K.~E. Torrance, ``Microfacet models for
  refraction through rough surfaces,'' in \emph{Rendering Techniques}, 2007.

\bibitem{Wagen2020}
J.-F. Wagen, ``Diffuse scattering and specular reflection from facets of
  arbitrary size and roughness using the computer graphics {GGX} model,'' in
  \emph{2020 International Symposium on Networks, Computers and Communications
  (ISNCC)}, 2020, pp. 1--5.

\bibitem{Saillard2011}
M.~Saillard and G.~Soriano, ``Rough surface scattering at low-grazing
  incidence: A dedicated model,'' \emph{Radio Science}, vol.~46, no.~5, 2011.

\bibitem{Priebe2011}
S.~Priebe, M.~Jacob, C.~Jansen, and T.~K\"urner, ``Non-specular scattering
  modeling for {THz} propagation simulations,'' in \emph{Proceedings of the 5th
  European Conference on Antennas and Propagation (EUCAP)}, 2011, pp. 1--5.

\bibitem{abramowitz-stegun}
M.~{Abramowitz} and I.~A. {Stegun}, \emph{Handbook of Mathematical Functions
  with Formulas, Graphs, and Mathematical Tables}.\hskip 1em plus 0.5em minus
  0.4em\relax New York City: Dover, 1964.

\end{thebibliography}

\end{document}